\documentclass[a4paper,12pt]{article} 
\usepackage[utf8]{inputenc}       
\usepackage[english]{babel} 
\usepackage{amssymb,amsmath,latexsym,enumerate}
\usepackage{cite}             
\usepackage{graphicx}
\usepackage{indentfirst}      
\usepackage{array,longtable,lscape}
\usepackage{color}
\usepackage[dvipsnames]{xcolor}
\numberwithin{equation}{section} 

\usepackage{geometry} 
\newcommand{\arsh}{\mathop{\rm arsinh}\nolimits}
\begin{document}
	
	\title{Free electron gas and electron-positron pair equilibrium in magnetic field}
	
	\author{G.S. Bisnovatyi-Kogan$^{1,2,3}$\footnote{gkogan@iki.rssi.ru}, I.A. Kondratyev$^{1,4}$\footnote{mrkondratyev95@gmail.com}\\
		{\it $^1$Space research institute RAS}\\
		{\it Profsoyuznaya st. 84/32, Moscow 117997} \\
		{\it $^2$National research nuclear university MEPhI}\\
		{\it Kashirskoye higway 31, Moscow 115409} \\
		{\it $^3$Moscow institute of physics and technology (State university)}\\
		{\it 141701 Moscow region, Dolgoprudny, Institutsky per. 9} \\
		{\it $^4$National reserach university} \\
		{\it $"$Higher school of economics$"$, Physics dept.}\\
		{\it Staraya Basmannaya st. 21/4k5, Moscow 105066}
	}
	\date{}
	\maketitle
	
	\begin{abstract}
		The thermodynamic properties of electron gas under the extreme conditions of high temperature, high matter density, and/or a strong magnetic field largely determine the behaviour of matter in upper layers of neutron stars and accretion columns of magnetized neutron stars in binary systems. A strong magnetic field in these objects makes the motion of electrons across the field essentially quantum. The possible electron degeneracy and relativism of electrons are also important. When studying accretion onto a magnetar in a binary system, the intensive generation of electron-positron pairs in the quantizing magnetic field should also be taken into account. We consider in detail the thermodynamic properties of a gas of free electrons in strong magnetic fields, taking into account their relativism and degeneracy, as well as the equilibrium creation of electron-positron pairs in a high-temperature plasma in the presence of a quantizing magnetic field.
	\end{abstract}

	\newpage
	\tableofcontents
	\section{Introduction}
	We start by briefly reviewing studies of the thermodynamical properties of free electrons in a magnetic field primarily related to astrophysical applications.
	In the 1920s, the development of quantum ideas and the discovery of spin by G Uhlenbeck and S Goudsmit \cite{ulga1925}, as well as the interpretation of spin as a separate quantum number corresponding to the proper particle’s angular momentum, led to the appearance of quantum statistics. The statistical properties of fermions taking into account the Pauli exclusion principle \cite{excprin} were first investigated by E Fermi and P Dirac \cite{fermi26, dirac26} in application to electrons (see also \cite{zommerfeld}). The statistical properties of bosons were studied by S Bose \cite{bose} for a photon gas and generalized by A Einstein for an ideal monoatomic gas \cite{einstein}. Later on, W Pauli proved the fundamental theorem relating the spin to statistics \cite{spinstat}. The theorem states that particles with half-integer spin satisfy the Fermi-Dirac statistic, while particles with integer spin satisfy the Bose- Einstein statistic. Quantum-statistical models of matter turned out to be very fruitful and explained many effects in physics and astronomy related to the behavior of electrons in both condensed matter in laboratories and stellar interiors.
	The Fermi-Dirac statistic application in research on compact objects in astrophysics started at the end of the 1920s in papers by R Fouler and Ya Frenkel \cite{fowler26,frenkel28}. Based on the quantum statistics, the equation of state of an electron nonrelativistic degenerate gas was derived in these papers. The equation of state of a degenerate gas with arbitrary relativism was obtained and first applied to construct models of white dwarfs with homogeneous density by E Stoner \cite{stoner30} (see also \cite{thomas}). The exact solution of the equilibrium equation of white dwarfs was first derived by S Chandrasekhar \cite{chandra31,chandra32} and L Landau \cite{landau32}. In these papers, an upper limit of the white dwarf mass was correctly found, which was earlier roughly estimated by Stoner (see paper \cite{yak94} for the history of these discoveries).
	
	Simultaneously, the development of the quantum theory of electron gas in external fields occurred. The classical theory of gas in a magnetic field suggests that thermodynamic functions of the gas remain the same as in the absence of the field—the so-called Bohr-van Leeuwen theorem (see Bohr’s paper \cite{bohr} written in 1911 and paper \cite{vanleuven}). This is due to the conservation of the particles’ Hamiltonian form after substituting a generalized momentum instead of an ordinary one. The statistical sum remains the same as in the absence of the field.
	The Schrodinger equation for a particle with spin 1/2 in a magnetic field was solved by Pauli \cite{pauli271}. At the same time, based on this solution, Pauli explained the paramagnetic properties of an electron gas in metals \cite{pauli272} via the interaction of the electron spin with the magnetic field, giving additional energy to electrons. Later, L D Landau \cite{landau30} showed that the motion of an electron across a magnetic field is also quantized, even ignoring the spin. The transversal (rotational) motion of the electron occurs in quantum orbits (the Landau quantization) with the transversal motion energy $E = (n_B + \frac{1}{2})\frac{\hbar eB}{m_e c}$. Here, $n_B$ is the Landau level’s number, $\hbar$ is the reduced Planck constant, $B$ is the magnetic field induction, $m_e$ is the electron mass, $e$ is the electron charge, and $c$ is the speed of light.
	
	If the transversal motion energy in the magnetic field is comparable to the characteristic kinetic energy of gas particles, the field, called quantizing, can affect the macroscopic properties of matter. The quantization of the orbital motion of electrons in a magnetic field explains the diamagnetic properties of the electron gas in metals. However, in general, the electron gas in metals is paramagnetic, because the paramagnetic part of the magnetic susceptibility is three times as high as the orbital diamagnetic part (see also \cite{rumer, LL5}). When electrons in the quantizing magnetic field become degenerate, a change in the field causes oscillations of the thermodynamic properties due to electron transitions between the Landau levels and their population redistribution.
	Quantum oscillations in strongly degenerate electrons in the quantizing field are the reason for the de Haas- van Alphen effect \cite{oscill2} (oscillations of the magnetization (magnetic moment) in metals \cite{LL5}) and the Shubnikov- de Haas effect (\cite{oscill1}; see also \cite{brch82}) — oscillations of the magnetoresistance observed in metals and semiconductors. Note that all thermodynamic functions of electron gas demonstrate such oscillations in the quantizing field \cite{shulman74}, with derivatives of the thermodynamic quantities oscillating more strongly \cite{rumer}. The kinetic coefficients of electrons also oscillate with a changing field \cite{potekhin1996}.
	
	Magnetic fields in white dwarfs can reach $\sim 10^6-10^8$ G on the surface and are too weak to alter the degenerate matter’s thermodynamic functions significantly. They appear mainly in the Zeeman effect near the stellar surface \cite{WDbook} (see, however, paper \cite{osthar68}, which assumes much stronger magnetic fields of white dwarfs). A sufficiently strong magnetic field of a white dwarf can substantially change (increase) its stability limit \cite{wd2014,mwd2015}. The super-Chandrasekhar mass of a white dwarf $M_{Ch} = 1.4 M_\odot$ ($M_\odot$ is the solar mass) may be responsible for the appearance of superluminous type I supernovae \cite{howell2006}.
	
	A more intensive investigation of the matter properties in strong magnetic fields started after the observational discovery of neutron stars with strong magnetic fields—radio pulsars \cite{hewish68} in 1967 and X-ray pulsars in binary systems \cite{xraypulsar} in 1971. Neutron stars have a radius of $R \sim 10 km$ and the characteristic mass $M \sim 1.4 M_\odot$. They are one of the end products of the evolution of massive stars after gravitational collapse. Neutron stars include a liquid core with a density exceeding the atomic nucleus density, the inner crust consisting of nuclei super-rich in neutrons, electrons, and free neutrons, and the outer crust composed of plasma of degenerate electrons and nondegenerate nonrelativistic nuclei (see, for example, [36, Ch. 1; 37, Ch. 8]). The strong degeneracy makes electrons in the outer crust an almost ideal gas, determining its thermodynamic properties.
	Magnetic fields of neutron stars reach incredibly high values inaccessible in laboratories. Estimates inferred from the observed rotational energy loss rate in the magneto-dipole model \cite{OstrGunn}, as well as the estimation of the magnetic field of a star after collapse with the magnetic flux conservation \cite{ginzburg70}, suggest that the magnetic field in radio pulsars is $\sim 10^{12}-10^{13} G$. It can be as high  $\sim 10^{14}-10^{15} G$ in a small group of neutron stars called magnetars \cite{magnetar, mazets1,mazets1,batse} (see also review \cite{bk2017} and references therein). Under such conditions, the magnetic field in the outer crust of a neutron star becomes quantizing for electrons and affects their thermodynamic properties.
	
	The first studies of the thermodynamics of electron gas in a quantizing field were carried out at the end of the 1960s in a series of papers \cite{canuto1968, canuto1968b, canuto1968c,canuto1969, caven77} in which expressions for the thermodynamic functions of a Fermi gas of noninteracting electrons in a magnetic field were obtained, and limiting transitions to the classical (nonquantizing) case were considered. The magnetic susceptibility of the neutron star crust was calculated in paper \cite{blanhern82}. In papers \cite{abrshap1991, laishap91, lai2001}, the Fermi-Dirac distribution taking into account the magnetic field in the strong degeneracy limit was applied to derive the equation of state of matter in the outer neutron star layers, and the strong field effects on beta-processes in the crust were considered. Similar research was performed in papers by G A Shul’man and co-authors \cite{shulman74,shulman93,shulman94}. Several studies were devoted to deriving thermodynamic functions of fermions heavier than electrons \cite{brod2000,ferrer} endowed with an anomalous magnetic moment \cite{strick}, with or without an electric charge. This could be important for magnetars in which the strong magnetic field effects can appear for baryons as well. Note, however, that the anomalous magnetic moment of an electron is small even in very strong fields \cite{duncan}. It can hardly significantly affect the properties of matter in neutron stars whose magnetic field cannot considerably exceed  $\sim 10^{18}$ G, according to Chandrasekhar and Fermi \cite{chanfer53}.
	
	The study of electron-positron pair creation is especially important for astrophysical applications dealing with a hot ($k_B T\sim m_ec^2$, where $k_B$ is the Boltzmann constant), low- density plasma (see, for example, \cite{bkzs72,svensson}). The understanding of the properties of a plasma, in which the pair number density is significantly higher than the baryon number density, is essential for the investigation of hot gas ejections (jets) from active galactic nuclei and blazars (see [63, Ch. 7]). Some thermodynamic properties of an equilibrium plasma without a magnetic field with possible electron-positron pair creation at high temperatures were obtained in papers \cite{pinaeva64,imshnad65,bkkazh66} in application to the stability of massive stars at the late stages of their evolution. The equation of state of electron-positron gas at high temperatures and densities was derived for a wide range of thermodynamic parameters in \cite{nad, nad742} (see also paper \cite{eegas1996}, summarizing analytical approximations of the Fermi-Dirac integrals for the arbitrary degeneracy degree and relativism of particles).
	
	The statistical properties of electrons and positrons in strong electromagnetic fields were first considered in the 1930s by Heisenberg and Euler \cite{heisen36} when studying pair creation from a vacuum. The equilibrium number densities of electrons and positrons in a hot plasma in a quantizing magnetic field were first calculated in papers \cite{shulman77,ishulman78,shulsekr77} (see also \cite{vsm89}). In the upper layers and atmospheres of neutron stars, of importance are the synchrotron radiation of the electron- positron pairs in the magnetic field \cite{bag2002} and photon emissions during spontaneous transitions between the Landau levels \cite{mp87} (see also [77, Ch. 10]), which is essential for pulsars and cosmic gamma-ray bursts. One of the radiation channels from neutron stars is due to the annihilation of an electron- positron pair into a neutrino-antineutrino pair in the magnetic field. The thermodynamic functions of the equilibrium electron-positron gas and neutrino luminosities are calculated in papers \cite{kamyak1992, kamyak1993}. The properties of electrons in strong magnetic fields are also important for studies of neutron star atmospheres (see, for example, paper \cite{vanriper} and review \cite{hardinglai}).
	
	Paper \cite{mush18} considered the creation of electron-positron pairs in the accretion column of a strongly magnetized neutron star in sources of ultraluminous X-rays (ULXs). Presently, the nature of ULXs in other galaxies remains unclear. It was assumed that these sources could be related to intermediate-mass black holes, $\sim 10^2\, - \, 10^4\, M_\odot$  \cite{ppdh2006}, or super-Eddington radiation from an accreting stellar-mass black hole. Accretion disks, in this case, are radiation- dominated and can reach a luminosity of $10^{40}$ erg s$^{-1}$ \cite{b2002,prh2003,fva2018}. Quite unexpectedly, super-Eddington X-ray pulsars had been discovered in several ULX sources \cite{bhw2014,cj2018}, suggesting their nature as strongly magnetized neutron stars.
	Accretion onto a strongly magnetized neutron star occurs along magnetic field lines \cite{bkf1969}, forming an accretion column \cite{basko76} in which the plasma temperature can attain $\sim 10^9$K. At such temperatures, the creation of electron-positron pairs is essential \cite{fh1964}. This should be taken into account in studies of the structure and observational appearance of the accretion column. The strong magnetic field changes both the pair creation rate and plasma radiative properties \cite{mush18, duncan}.
	
	In the present paper, we consider the thermodynamic properties of an electron gas in magnetic fields in detail, taking into account the degeneracy and relativism of electrons in the presence of quantizing and supercritical magnetic fields. We study how the thermodynamic properties change when transiting from nonquantizing to quantizing fields in a nonrelativistic plasma and for relativistic electrons, for which the quantizing field must be supercritical.
	The properties of an electron plasma in a quantizing magnetic field for degenerate and nondegenerate cases are considered. We show that, in a nondegenerate plasma, the thermodynamic properties smoothly and monotonically change with increasing magnetic field. Upon increasing the degeneracy degree, maxima and minima, corresponding to magnetic oscillations of the thermodynamic functions, appear in this dependence. In the approximation of fully degenerate electrons, sharp peaks arise, transforming into smooth dependences after the magnetic field has exceeded the value at which all electrons are at the lowest Landau level \cite{laishap91,lai2001}. We have considered pair creation in a high-temperature plasma at thermodynamic equilibrium in the presence of a quantizing magnetic field with a possible strength exceeding the critical value $B_{cr} = m_e^2c^3/e\hbar = 4.414\cdot10^{13}$ G. The presence of a magnetic field decreases the temperature at which electron-positron pairs can be effectively generated. On the other hand, the number density of electron-positron pairs in the quantizing magnetic field increases with the magnetic field under a constant temperature and chemical potential of electrons \cite{shulman77,kamyak1992,mush18}.
	Note also textbooks on statistical mechanics that consider in detail some of the topics in the present paper related to the properties of an electron gas in a magnetic field, including a quantizing one. These include books by R Kubo [92, Ch. 4] for nonrelativistic particles and R Hakim [93, Ch. 12], in which relativistic fermions are also treated.

	\section{Landau quantization: discrete levels of electrons in a magnetic field}
	
	From the point of view of classical physics, an electron moves in a circular orbit across a magnetic field with angular frequency $\omega_B = \frac{eB}{m_ec}$ and orbital radius $R_L = \frac{p_\perp c}{eB}$, where $p_\perp$ is the electron’s transversal momentum.
	The energy of quantum $E_B$ corresponding to the electron cyclotron rotational frequency $\omega_B$ in the magnetic field is
	
	\begin{equation}
	E_{B}=\hbar \omega_B= \hbar \frac{eB}{m_e c}.
	\label{en0}
	\end{equation}
	$E_{B}$ reaches the electron rest-mass energy $m_e c^2$ for the magnetic field $B=B_{cr}$
	
	\begin{equation}
	B_{cr}= \frac{m_e^2 c^3}{e \hbar }=4.414\cdot10^{13}\,\,{\mbox{G}},
	\label{enb}
	\end{equation}
	which is called the critical or Schwinger magnetic field \cite{bag2002}. For $B>B_{cr}$, the electron in the magnetic field becomes relativistic. In quantum mechanics, finite motion across the magnetic field is quantized. The energy levels of a nonrelativistic electron in a weak, $B \ll B_{cr}$, homogeneous field are determined by the formula \cite{LL3}
	
	\begin{equation}
	E_n = \frac{p_z^2}{2m_e} + \frac{2n_B + 1 + \sigma_s}{2}\hbar \omega_B,
	\label{en}
	\end{equation}
	where $p_z = p_\parallel$ is the electron’s momentum along the field (the z-axis), the integer positive number nB defines the Landau level number, and $\sigma_s = \pm 1$ corresponds to the spin projection. After having numbered the Landau levels by integer numbers $n$, we obtain
	
	\begin{equation}
	E_n = \frac{p_z^2}{2m_e} + n \hbar \omega_B=\frac{p_z^2}{2m_e} +
	n \frac{eB \hbar}{m_e c}, \quad n= \frac{2n_B + 1 + \sigma_s}{2}.
	\label{energy}
	\end{equation}
	Here, the degeneracy of the nth level (the statistical weight of the level) is $g_n = 2$ for $n > 0$ and $g_n = 1$ for $n = 0$. The square of the transversal momentum can be expressed as \cite{bag2002}
	
	\begin{equation}
	p^2_{\perp,n} = 2\frac{eB\hbar}{c}(n -\frac{\sigma_s}{2})= 2\frac{eB\hbar}{c}(n_B +\frac{1}{2}),
	\label{tr_mom}
	\end{equation}
	That is, in the ground state with $n=n_B=0,\,\, \sigma_s=-1$, and zero rotational energy, the electron has a nonzero transversal momentum. The zero energy of the ground level is related to the negative contribution to the zero-state energy in (\ref{energy}) of the spin term with $\sigma_s=-1$, whereas the spin term does not contribute to the orbital angular momentum. 
	
	In the relativistic case, where the electron energy is comparable to or higher than the rest-mass energy, formula \eqref{energy} should be substituted with its relativistic extension \cite{LL4} following from the solution of Dirac equation \cite{sokolovternov} for a particle in an external electromagnetic field:
	
	\begin{equation}
	E_n(p_z) = \sqrt{m_e^2c^4 + p_z^2c^2 + 2n\hbar e B c} = m_ec^2\sqrt{1 + p^2 + 2bn}.
	\label{relen}
	\end{equation}
	
	Here, $b = B/B_{cr}$ (see \eqref{enb}), and $p_z$ is the electron’s momentum along the field, $p = p_z/m_ec$. Expression \eqref{relen} is applicable to arbitrary magnetic field induction and the electron’s longitudinal momentum. The wave functions of the electron in an external magnetic field are not used below in the explicit form; the solutions of the Dirac and Schrodinger equations are considered in \cite{LL4,LL3}, respectively.

	\section{ Nonrelativistic electron Fermi gas and $e^+e^-$-pairs in subcritical magnetic field $B \ll B_{cr}$ }
	
	\subsection{Fermi-Dirac distribution of nonrelativistic electrons in a magnetic field}
	
	To obtain thermodynamic functions of an electron gas in a magnetic field, it is necessary to sum over all quantum states with the Fermi-Dirac distribution function. The phase ‘volume’ (surface) in the transversal plane of the $n$th Landau level $\Delta V_{ph}^{(n)}$ in the axially symmetric case is determined by the difference between transversal momentum squares of the neighboring levels. Taking into account (\ref{tr_mom}), we have
	
	\begin{equation}
	\Delta V_{ph}^{(n)}=\pi(p^2_{\perp,(n+1)}-p^2_{\perp,n}) = 2\pi\frac{eB\hbar}{c}.
	\label{phv}
	\end{equation}
	
	\noindent The number density of electrons $n_{en}$ at the $n$th Landau level can be obtained by integrating the Fermi-Dirac distribution with the $n$th level transversal motion energy from (\ref{energy}) over the longitudinal momentum $p_z$ along the field:

	\begin{equation}
	f_n=\frac {1}{1+\exp[(E_n-\mu_e)/k_B T]}=
	\frac {1}{1+\exp[(\frac{p_z^2}{2m_e} + n \frac{eB \hbar}{m_e c}-\mu_e)/k_B T]}.
	\label{fd}
	\end{equation}
	Here, $\mu_e$ is the chemical potential of electrons, and $T$ is the electron gas temperature. As a result, taking into account the phase volume (\ref{phv}) and the phase volume of the elementary cell $(2\pi\hbar)^3$, we get
	
	\begin{equation}
	n_{en}=
	\frac{\Delta V_{ph}^{(n)}}{(2\pi\hbar)^3}g_n\int_{-\infty}^{+\infty} f_n dp_z
	=\frac{1}{4\pi^2} \frac{eB}{c\hbar^2}g_n\int_{-\infty}^{+\infty} f_n dp_z.
	\label{nn}
	\end{equation} 
	The total concentration of the magnetized electron Fermi gas $n_e$ is found by summing (\ref{nn}) over all levels. This results in \cite{canuto1968,shulman77,shulman93,kubo,lai2001}
	
	\begin{equation}
	\begin{gathered}
	n_e = \sum_{n = 0}^{\infty} n_{en}=
	\frac{1}{4\pi^2} \frac{eB}{c\hbar^2}\sum_{n = 0}^{\infty}g_n\int_{-\infty}^{+\infty} f_n dp_z.
	\end{gathered}
	\label{relconsentr}
	\end{equation}
	With decreasing magnetic field, the electrons in equilibrium fill up an increasingly large number of Landau levels. Here, the interlevel distance decreases, which allows a quasi-classical treatment. In the limit of large $n$, it becomes possible to consider it a continuous value and pass in (\ref{relconsentr}) from summation to integration over $n$. We then obtain
	
	\begin{equation}
	\begin{gathered}
	n_e =
	\frac{1}{2\pi^2} \frac{eB}{c\hbar^2}\int_0^{\infty}dn \int_{-\infty}^{+\infty} f_n dp_z.
	\end{gathered}
	\label{nk}
	\end{equation}
	Here, we have adopted $g_n=2$ for all levels, because with decreasing magnetic field, the zero-level contribution also decreases, and the error due to doubling coefficient $g_0=1$ at the zero term becomes negligible. Let us change variables in (\ref{fd}):
	
	\begin{equation}
	n\frac{eB\hbar}{m_e c}=\frac{p_\perp^2}{2m_e}.
	\label{dnk}
	\end{equation}
	Then, (\ref{nk}) takes the form
	
	\begin{equation}
	\begin{gathered}
	n_{ed} =
	\frac{1}{4\pi^2 \hbar^3}\int_0^{\infty}dp_\perp^2 \int_{-\infty}^{+\infty} \frac {1}{1+\exp[(\frac{p_z^2}{2m_e} + \frac{p_\perp^2}{2 m_e}-\mu_e)/k_B T]} dp_z.
	\end{gathered}
	\label{nk1}
	\end{equation}
	In (\ref{nk1}), it is convenient to change the cylindrical coordinates $(p_\perp,\,\, \phi,\,\, p_z)$ into spherical ones $(p,\,\,\theta,\,\,\phi)$ in the phase space due to the phase space isotropy in the absence of the magnetic field. In the spherical coordinates, (\ref{nk1}) takes the following form:
	
	\begingroup
	\small
	\begin{equation}
	\begin{gathered}
	n_{ed} =
	\frac{1}{\pi^2 \hbar^3}\int_{0}^{\infty} \frac {p^2 dp}{1+\exp(\frac{p^2}{2m_e k_B T}-\frac{\mu_e}{k_B T})}
	=\frac{\sqrt 2}{\pi^2}\left(\frac{m_e k_B T}{\hbar^2}\right)^{3/2}\int_{0}^{\infty} \frac {\sqrt u du}{1+\exp(u-\frac{\mu_e}{k_B T})}\\
	=\frac{\sqrt 2}{\pi^2}\left(\frac{m_e k_B T}{\hbar^2}\right)^{3/2}F_{1/2}(\frac{\mu_e}{k_B T}) .
	\end{gathered}
	\label{nk0}
	\end{equation}
	\endgroup
	which exactly coincides with the expression for the electron number density without the field \cite{LL5,bisn}. In formula \eqref{nk0}, we have made the substitution $u = p^2/(2m_e k_B T)$. We have also used the standard notation of the Fermi integrals [36]:
	
	\begin{equation}
	F_\nu(\xi)= \int_{0}^{\infty} \frac{u^{\nu} du}{1+\exp(u-\xi)}.
	\label{nkf}
	\end{equation}
	With increasing magnetic field, the accuracy of the integral representation (\ref{nk1}) decreases. To reach sufficient precision, one should calculate terms from the series in (\ref{relconsentr}) and (\ref{fd}).
	
	\subsection{Nondegenerate nonrelativistic electron gas in a magnetic field}
	
	For a nondegenerate gas, instead of the Fermi function (\ref{fd}), one can use the Maxwell-Boltzmann distribution,
	
	\begin{equation}
	f_{nmb}=\exp[-(E_n-\mu_e)/k_B T]=
	\exp[-(\frac{p_z^2}{2m_e} + n \frac{eB \hbar}{m_e c}-\mu_e)/k_B T],
	\label{fdmb}
	\end{equation}
	This enables us to obtain an analytical expression for the sum of a series which, unlike (\ref{relconsentr}) with $f_n$ from (\ref{fd}), can be written as
	
	\begin{eqnarray}
	n_e =
	\frac{1}{4\pi^2} \frac{eB}{c\hbar^2}\sum_{n = 0}^{\infty}g_n\int_{-\infty}^{+\infty} \exp[-(\frac{p_z^2}{2m_e} + n \frac{eB \hbar}{m_e c}-\mu_e)/k_B T] dp_z \nonumber\\
	=\frac{1}{4\pi^2} \frac{eB}{c\hbar^2}\sum_{n = 0}^{\infty}g_n \exp(-n \frac{eB \hbar}{m_ec k_B T})\times \int_{-\infty}^{+\infty} \exp[-(\frac{p_z^2}{2m_e} -\mu_e)/k_B T] dp_z.
	\label{nmd}
	\end{eqnarray}
	Here, the series in the first term is calculated precisely analytically as the sum of a geometric progression. We have
	
	\begin{eqnarray}
	\frac{1}{4\pi^2} \frac{eB}{c\hbar^2}\sum_{n = 0}^{\infty}g_n \exp(-n \frac{eB \hbar}{m_ec k_B T})
	=\frac{1}{2\pi^2} \frac{eB}{c\hbar^2}\{\left[1-\exp(-\frac{eB \hbar}{m_ec k_B T})\right]^{-1}-\frac{1}{2}\}
	\label{nmd1}
	\end{eqnarray}
	where  $g_0=1$, $g_n=2$ for $n\ge 1$ are taken into account. After substituting (\ref{nmd1}) into (\ref{nmd}), we arrive at
	\begin{eqnarray}
	n_e
	=\frac{1}{2\pi^2} \frac{eB}{c\hbar^2}\{\left[1-\exp(-\frac{eB \hbar}{m_ec k_B T})\right]^{-1}-\frac{1}{2}\}
	\times \int_{-\infty}^{+\infty} \exp[-(\frac{p_z^2}{2m_e} -\mu_e)/k_B T] dp_z,
	\label{nmd2}
	\end{eqnarray}
	see also \cite{shulman77,kamyak1992}.
	Formula \eqref{nmd2} is an exact expression for the electron number density in a nonrelativistic, nondegenerate gas for an arbitrary magnetic field, ranging from a weak field in the quasi-classical limit to a strongly quantizing field below the critical value, $B << B_{cr}$ from (\ref{enb}). In the limiting case of a weak nonquantizing field, one can expand in the first term the exponent in series and ignore there the factor 1/2 due to its relative smallness. We then obtain
	
	\begin{eqnarray}
	n_{e0}
	=\frac{m_e k_B T}{2\pi^2\hbar^3}
	\int_{-\infty}^{+\infty} \exp[-(\frac{p_z^2}{2m_e} -\mu_e)/k_B T] dp_z.
	\label{nmd3}
	\end{eqnarray}
	In this expression, only integration over $p_z$ remains, because integration over $p_\perp^2$ has already been performed when summing the series. Indeed, by integrating expression (\ref{nk1}) for the nondegenerate gas over the transversal momentum, we obtain the formula coinciding with (\ref{nmd3}):
	
	\begin{eqnarray}
	n_{e0} =
	\frac{1}{4\pi^2 \hbar^3}\int_0^{\infty}dp_\perp^2 \int_{-\infty}^{+\infty} \exp[-(\frac{p_z^2}{2m_e} + \frac{p_\perp^2}{2 m_e}-\mu_e)/k_B T] dp_z \nonumber\\
	=\frac{m_e k_B T}{2\pi^2\hbar^3}
	\int_{-\infty}^{+\infty} \exp[-(\frac{p_z^2}{2m_e} -\mu_e)/k_B T] dp_z=2\left(\frac{m_e k_B T}{2\pi\hbar^2}\right)^{3/2} \exp(\frac{\mu_e}{k_B T}).
	\label{nk2}
	\end{eqnarray}
	Here, we have employed the values of the integrals
	
	\begin{eqnarray}
	\int_{-\infty}^{+\infty} \exp(-\frac{p_z^2}{2m_e k_B T}) dp_z = (2\pi m_e k_B T)^{1/2}, \,\,
	\int_0^{\infty}\exp(-\frac{p_\perp^2}{2 m_e k_B T}) dp_\perp^2=2 m_e k_B T.
	\label{nmd4}
	\end{eqnarray}
	In a nondegenerate, nonrelativistic electron gas, the electron number density for arbitrary quantizing and nonquantizing magnetic fields is determined by the same formula that follows from (\ref{nmd2}), (\ref{nmd4}):
	
	\begin{eqnarray}
	n_e=
	\frac{(2\pi m_e k_B T)^{1/2}}{2\pi^2} \frac{eB}{c\hbar^2}\{\left[1-\exp(-\frac{eB \hbar}{m_ec k_B T})\right]^{-1}-\frac{1}{2}\} \exp(\frac{\mu_e}{k_B T}).
	\label{nmd5}
	\end{eqnarray}
	In the limiting cases, we have
	
	\begin{eqnarray}
	n_e=\frac{1}{\sqrt 2\pi^{3/2}}
	\left(\frac{m_e k_B T}{\hbar^2}\right)^{3/2}\exp(\frac{\mu_e}{k_B T})\quad {\mbox{for a nonquantizing field at }}
	\quad \frac{\hbar \omega_B}{k_BT} \ll 1, \nonumber\\
	n_e=\frac{(2\pi m_e k_B T)^{1/2}}{4\pi^2}\frac{eB}{c\hbar^2}
	\exp(\frac{\mu_e}{k_B T})\quad {\mbox{for a quantizing field at }}
	\quad \frac{\hbar \omega_B}{k_BT} \gg 1. \qquad
	\label{nmd5a}
	\end{eqnarray}
	
	\subsection{Degenerate gas and magnetic oscillations}
	
	Taking into account (\ref{enb}), we can present the exponent power in (\ref{nmd5}) in the form
	
	\begin{equation}
	\frac{eB \hbar}{m_e c k_B T}=\frac{\hbar \omega_B}{k_BT}=\frac{B}{B_{cr}}\cdot \frac{m_e c^2}{k_B T}=\frac{B}{\gamma_T B_{cr}}, \qquad
	\gamma_T=\frac{k_B T}{m_e c^2}.
	\label{nmd6}
	\end{equation}
	In nonrelativistic plasma $ \gamma_T < 1$, $ B< B_{cr}$, and the magnetic field becomes quantizing for $B_{cr}> B>\gamma_T B_{cr}$, when the exponent powers in (\ref{nmd5}) and (\ref{fd}) become large, and the series terms in (\ref{relconsentr}) rapidly decrease. In the limit $B_{cr}> B\gg\gamma_T B_{cr}$, the first term in (\ref{relconsentr}) mainly contributes to the electron number density, which corresponds to almost all electrons being at the ground Landau level. Here, $\hbar \omega_B \gg k_BT$, and in formula \eqref{relconsentr} we can take into account only the first term:
	
	\begin{equation}
	n_e = \frac{1}{4\pi^2} \frac{eB}{c\hbar^2}\int_{-\infty}^{+\infty} \frac{dp_z}{1 +
		\exp[(\frac{p_z^2}{2m_e}-\mu_e)/k_B T]}.
	\label{strongBnonrel}
	\end{equation}
	Under these conditions, the electron gas behaves like a monoatomic gas of free particles with a one-dimensional Fermi distribution function, but with the number density increased by a factor proportional to the field. The divergence from a simple one-dimensional gas is due to electrons having the nonzero transversal momentum $p_{\perp,0} = \frac{eB\hbar}{c}$ (\ref{tr_mom}) even with zero rotational energy, which correspondingly increases the transverse phase volume and the corresponding coefficient in formula \eqref{strongBnonrel}.
	
\begin{figure}[!htp]
		\centering
		\includegraphics[width=9.0cm,height=6.0cm]{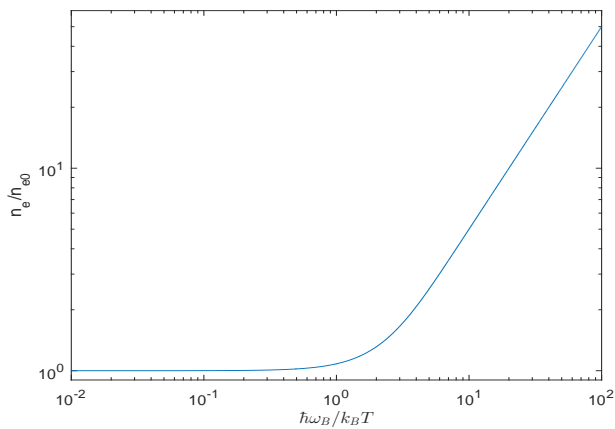}
		\caption{ Electron number density ratio with the same $\mu_e$ in the case of quantizing and nonquantizing fields $\frac{n_e}{n_{e0}}$ as a function of the parameter $\frac{\hbar \omega_B}{k_B T}$ in a nondegenerate, nonrelativistic gas. }
		\label{fig:fig1}
	\end{figure}

	Taking into account (\ref{nmd4}),(\ref{nmd6}) and (\ref{nk2}), from (\ref{nmd5}) we get
	
	\begin{eqnarray}
	\frac{n_e}{n_{e0}}=
	\frac{\hbar \omega_B}{k_B T} \{\left[1-\exp(-\frac{\hbar\omega_B}{k_BT})\right]^{-1}-\frac{1}{2}\}, \quad \frac{\hbar\omega_B}{m_e c^2}<1.
	\label{nmd7}
	\end{eqnarray}
	Thus, for the same chemical potential $\mu_e$, the dependence on the magnetic field relative to the equilibrium number density of nondegenerate, nonrelativistic electrons $\frac{n_e}{n_{e0}}$ reduces to the dependence on one parameter $\frac{\hbar \omega_B}{k_B T}$, shown in Fig.\ref{fig:fig1}
	
\begin{figure}[!htp]
		\centering
		\includegraphics[width=9.0cm,height=6.0cm]{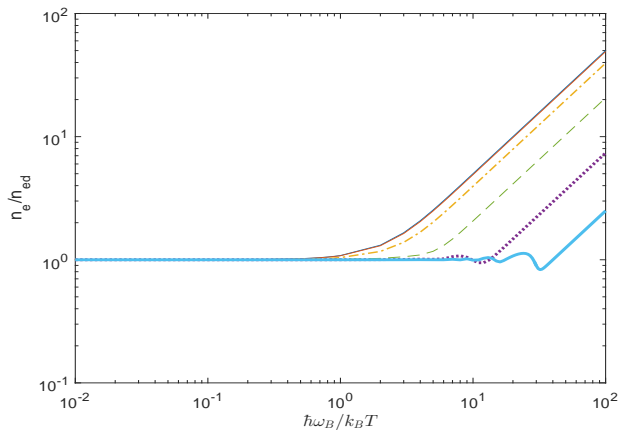}
		\caption{ Electron number density ratio in the case of quantizing and nonquantizing magnetic fields $\frac{n_e}{n_{ed}}$ as a function of the parameter $\frac{\hbar \omega_B}{k_B T}$ for different chemical potentials: $\frac{\mu_e}{k_B T}$ = —10, —3, 0, 3, 10, 30, respectively, for the blue, red (almost coincident) solid, yellow dashed-dotted, green dashed, violet dotted, and solid blue curves. The two upper curves almost coincide with the curve in Fig. 1 for a nondegenerate gas. The oscillations in the relative electron number density in the strong degenerate case are due to discrete electron transitions between the Landau levels with increasing quantizing magnetic field. }
		\label{fig:fig2}
	\end{figure}

	For nonrelativistic electrons with an arbitrary degeneracy degree, the relative electron number density depends on two parameters. Besides the dependence on the parameter $\frac{\hbar \omega_B}{k_B T}$, an explicit dependence on the general chemical potential, related to the electron degeneracy degree, expressed as $\frac{\mu_e}{k_B T}$, arises. From (\ref{fd}), (\ref{nk0}) and \eqref{relconsentr} and taking into account (\ref{nmd6}), we obtain the number density ratio in the form 
	
	\begin{equation}
	\begin{gathered}
	\frac{n_e}{n_{ed}} =\frac{\hbar \omega_B}{4k_B T} \sum_{n = 0}^{\infty} g_n\int_{-\infty}^{+\infty}
	\frac {dy}{1+\exp(y^2 + n \frac{\hbar\omega}{k_B T}-\frac{\mu_e}{k_B T})}/ F_{1/2}(\frac{\mu_e}{k_B T}),
	\end{gathered}
	\label{nmd8}
	\end{equation}
	Here, we have made the substitution $y^2=\frac{p_z^2}{2m_e k_B T}$. The dependence of the relative number density $\frac{n_e}{n_{e0}}$ on the parameter $\frac{\hbar \omega_B}{k_B T}$ for different $\frac{\mu_e}{k_B T}$ is presented in Fig. \ref{fig:fig2}.

\begin{figure}[!htp]
		\centering
		\includegraphics[width=9.0cm,height=6.0cm]{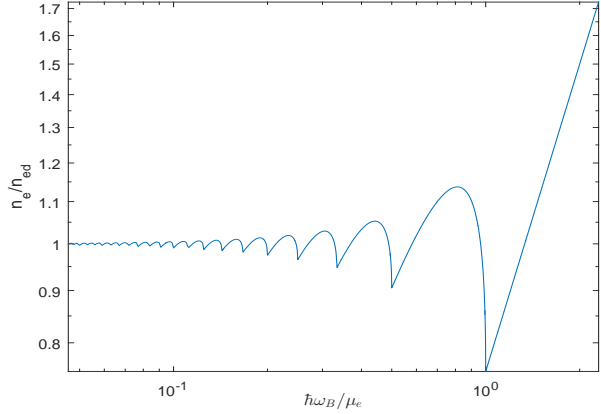}
		\caption{ Electron number density ratio in the case of quantizing and nonquantizing magnetic field $\frac{n_e}{n_{ed}}$ as a function of the parameter $\frac{\hbar \omega_B}{\mu_e}$ in the strongly degenerate case, as calculated by the first formula in \eqref{nmd9}. The temperature effects ‘smear’ the oscillations by making them smoother. The last peak in the number density oscillations occurs at $\frac{\hbar \omega_B}{\mu_e} = 1$ according to \eqref{nmd11}. }
		\label{fig:fig3}
	\end{figure}

\begin{figure}[!htp]
		\centering
		\includegraphics[width=9.0cm,height=6.0cm]{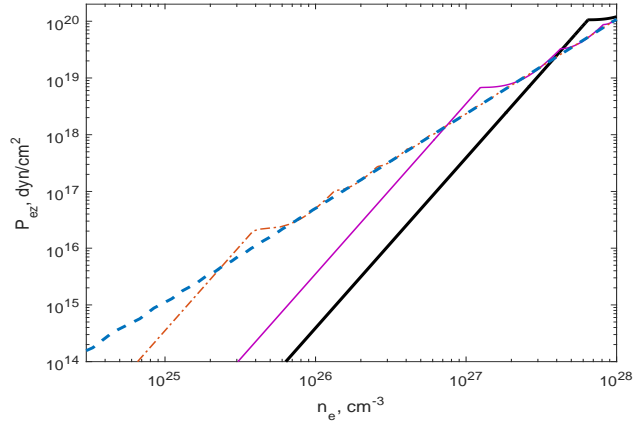}
		\caption{ Degenerate electron gas pressure as a function of the electron density in a magnetic field with strengths $B = 8\cdot10^9; 4\cdot10^{10}; 4\cdot10^{11}; 1.2\cdot10^{12}$  G (blue dashed, red dashed-dotted, magenta and black solid curves, respectively). The pressure dependence on the magnetic field is most pronounced only in the strongly degenerate case. For example, the blue dashed curve describes the gas in a relatively weak magnetic field, in which the electron pressure follows the adiabatic law with index $\gamma = 5/3$. As the field increases, an increasingly fewer number of Landau levels are occupied by electrons. In a sufficiently strong field (black curve, $B = 1.2\cdot10^{12}$ G), in the quantizing limit for $n_e \lesssim 6\cdot10^{27}$ cm-3, only the ground Landau level is populated, and the pressure is more sensitive to the number density, $P_{ez} \propto n_e^3$. When the electron density increases, the first excited Landau level starts becoming occupied. }
		\label{fig:fig4}
	\end{figure}

\begin{figure}[!htp]
		\centering
		\includegraphics[width=7.0cm,height=4.5cm]{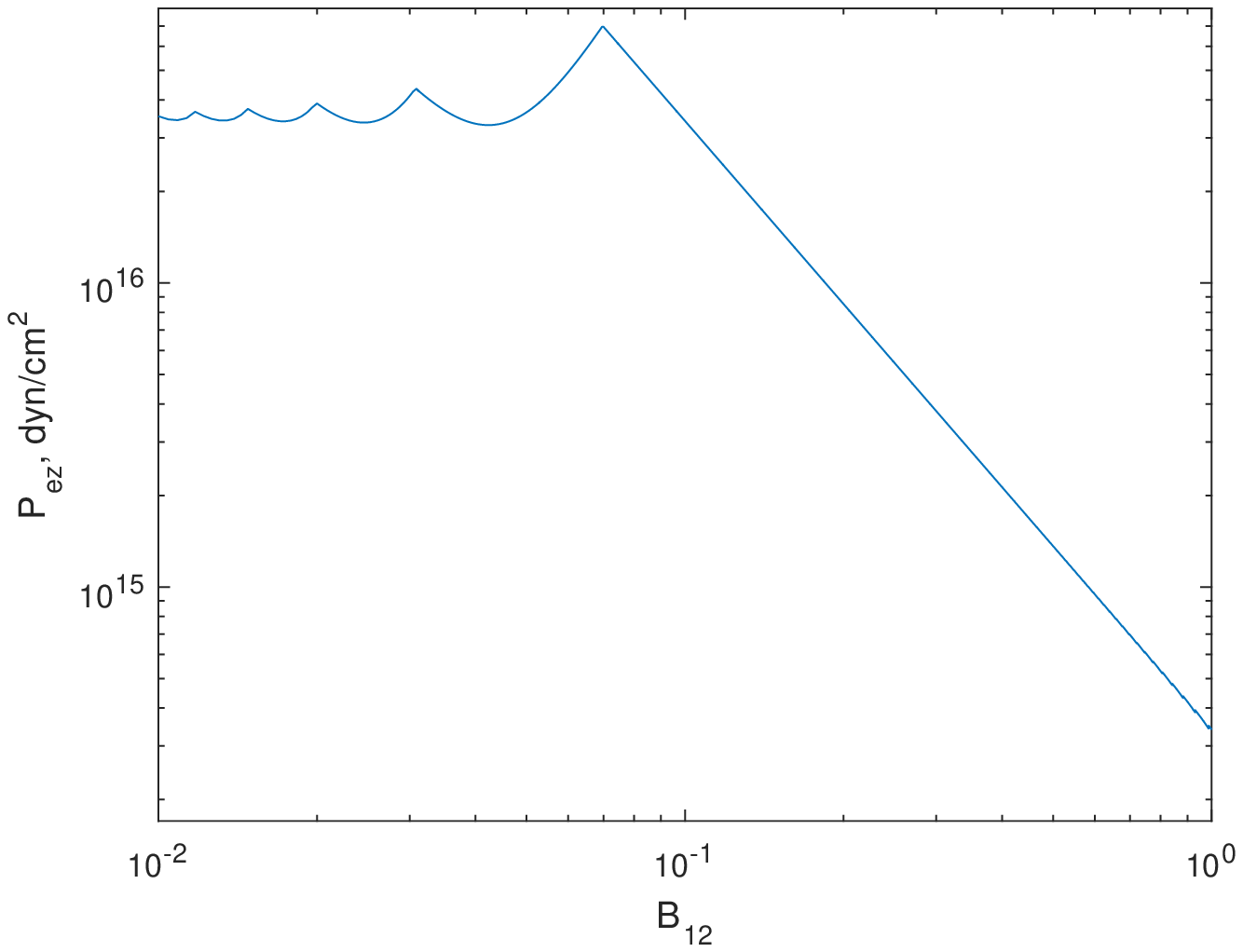}
		\includegraphics[width=7.0cm,height=4.5cm]{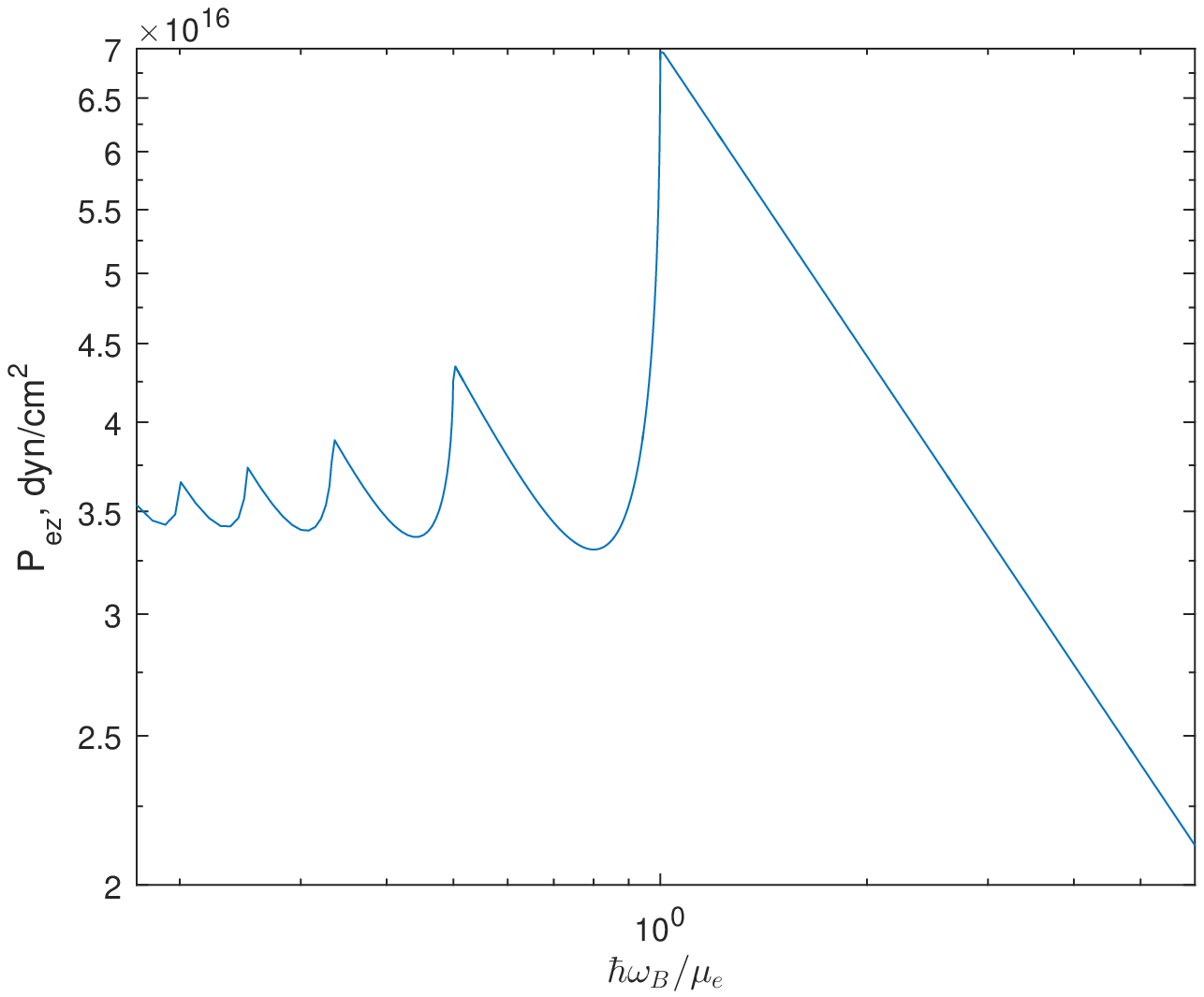}
		\caption{ (left) Degenerate electron gas pressure as a function of the magnetic field $B$ (in the units of $10^{12}$ G) for an electron density of $n_e = 8\cdot10^{25}$ cm$^{-3}$. (right) The same curve as in the left figure calculated near the transition from nonquantizing to strongly quantizing fields as a function of ratio $\hbar \omega_B/\mu_e$. The pressure oscillations are similar to the degenerate electron gas density oscillations in a quantizing magnetic field. With increasing magnetic field, an increasingly fewer number of Landau levels are occupied by electrons. After the last pressure maximum, all electrons transit to the ground Landau level, and $P_{ez} \propto 1/B^2$ according to \eqref{nmd13} and \eqref{nmd9}. }
		\label{fig:fig5}
	\end{figure}

\begin{figure}[!htp]
		\centering
		\includegraphics[width=7.0cm,height=4.5cm]{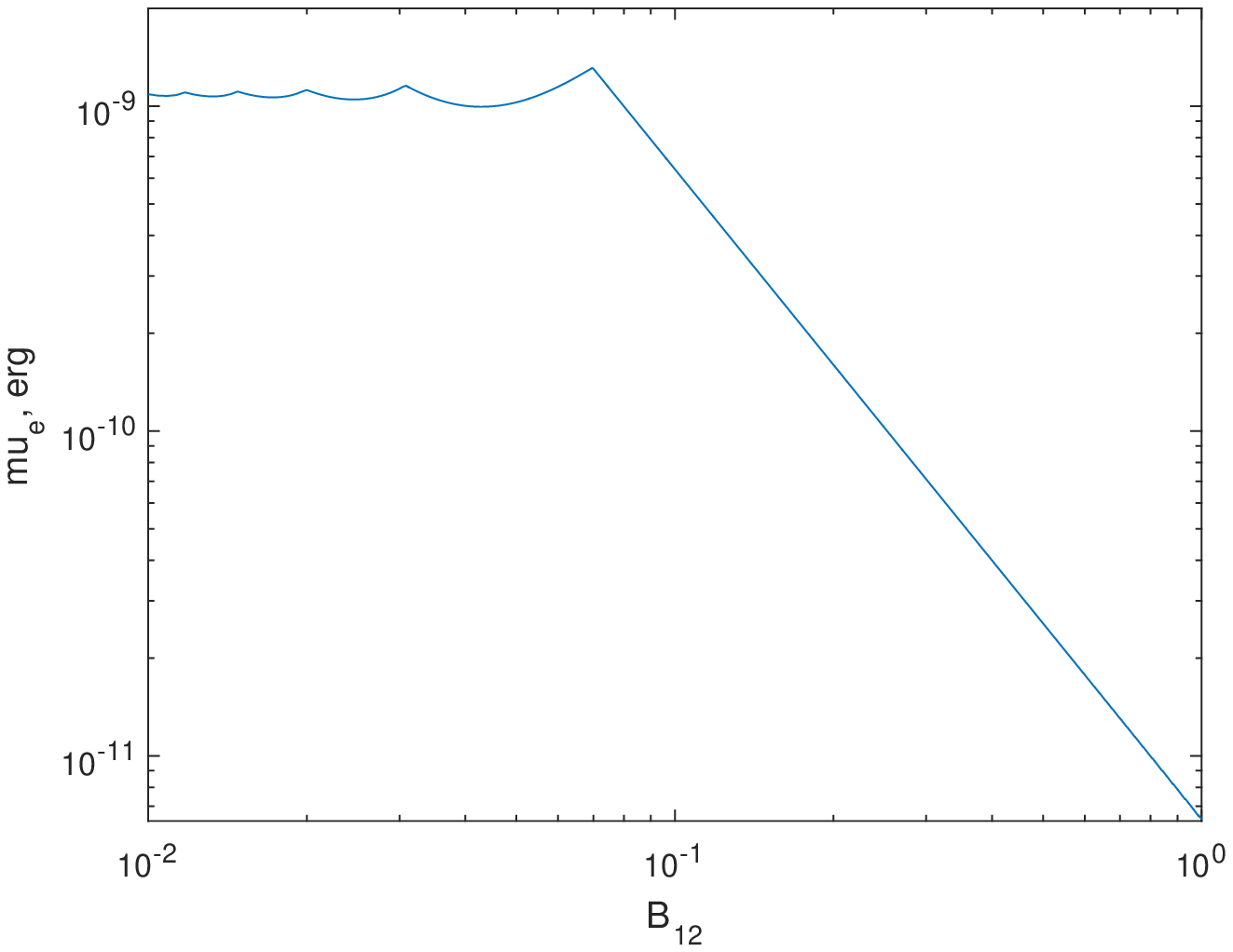}
		\includegraphics[width=7.0cm,height=4.5cm]{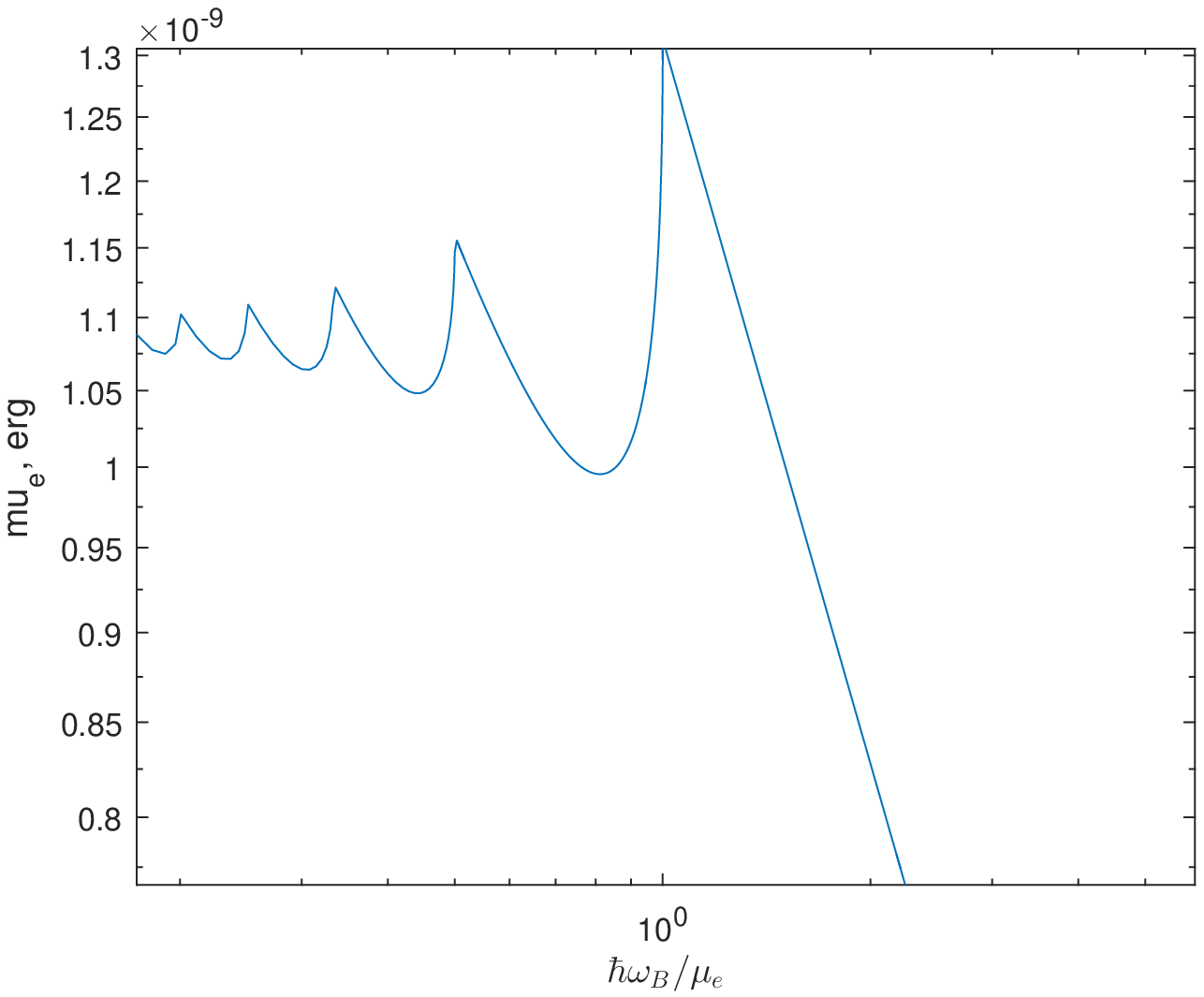}
		\caption{ (left) Chemical potential of an electron gas in a magnetic field calculated using \eqref{nmd9} in the strongly degenerate case as a function of the magnetic field $B$ (in the units of $10^{12}$ G) at an electron density of $n_e = 8\cdot10^{25}$ cm$^{-3}$. (right) The same curve as in the left figure calculated near the transition from nonquantizing to strongly quantizing fields as a function of ratio  $\hbar \omega_B/\mu_e$. Clearly, the jump in the derivative of the chemical potential occurs here when it reaches a maximum with increasing field and decreasing number of populated Landau levels. At the same time, the dependence of the relative number density on the same quantity, shown in Fig. 3, has a minimum at the derivative jump point. }
		\label{fig:chempot}
	\end{figure}

	From (\ref{nmd8}), in the limiting case of nondegenerate electrons, after summing up the series over $n$, we obtain (\ref{nmd7}). For strongly degenerate electrons, the integration is performed in the limits $(-\sqrt{\frac{\mu_e-n\hbar\omega_B}{k_B T}}, \,\, \sqrt{\frac{\mu_e-n\hbar\omega_B}{k_B T}})$ in integral (\ref{nmd8}) and in the limits $(0,\,\, \frac{\mu_e}{k_B T})$ in integral (\ref{nkf}). As a result, in the strongly degenerate case, we obtain the following relation:
	
	\begin{equation}
	\begin{gathered}
	\frac{n_e}{n_{ed}}=\frac{3\hbar \omega_B}{4\mu_e}\sum_{n = 0}^{\frac{\mu_e}{\hbar \omega_B}} g_n
	\sqrt{1-n\frac{\hbar \omega_B}{\mu_e}},\\ \int_0^{\mu_e/k_B T} y^{1/2} dy=
	\frac{2}{3}\left(\frac{\mu_e}{k_B T}\right)^{3/2},\quad n_{ed}=\frac{1}{3\pi^2\hbar^{3/2}}\bigg(\frac{2m_e}{\hbar}\bigg)^{3/2} \mu_e^{3/2}.
	\end{gathered}
	\label{nmd9}
	\end{equation}
	For $\hbar \omega_B>\mu_e$, all electrons occupy the ground Landau level. Then,
	
	\begin{equation}
	\frac{n_e}{n_{ed}}=\frac{3\hbar \omega_B}{4\mu_e}.
	\label{nmd10}
	\end{equation}
	The appearance of oscillations in the dependence of the function $\frac{n_e}{n_{ed}}$ on the magnetic field $B$ can be easily seen in the full degeneracy limit \eqref{nmd9} by writing up this function for the following arguments (for the occupation of levels with $n$ = 0,1):
	
	\begin{equation}
	\begin{gathered}
	\frac{n_e}{n_{ed}}=\frac{3\hbar \omega_B}{4\mu_e}(1+2\sqrt{1-\frac{\hbar \omega_B}{\mu_e}}),\quad {\mbox {for\,\,\,}} \mu_e/2\le\hbar \omega_B\le\mu_e,\\
	\frac{n_e}{n_{ed}}=\frac{3\hbar \omega_B}{4\mu_e}, \quad{\mbox {for\,\,\,}} \hbar \omega_B\ge\mu_e.
	\end{gathered}
	\label{nmd11}
	\end{equation}
	For $\frac{\hbar \omega_B}{\mu_e}=1$, where all electrons transit to the zero level, $\frac{n_e}{n_{ed}}= \frac{3}{4}$, corresponding to the right minimum in Fig. \ref{fig:fig3}. As follows from \eqref{nmd11}, in the fully degenerate case (see Fig. \ref{fig:fig3}), the function appears as a sharp angle with its vertex at the minimum point near this minimum. Several oscillations with a lower amplitude in Figs \ref{fig:fig2} and \ref{fig:fig3} are related to the possibility of transition at the higher Landau levels with decreasing magnetic field [22]. The dependences of the chemical potential $\mu_e$ and the longitudinal pressure $P_{ez}$ of degenerate electrons on the magnetic field are shown in Figs \ref{fig:fig4}-\ref{fig:chempot}. Oscillations of different physical quantities with the magnetic field in the quantizing field are due to discrete transitions of electrons between the Landau levels with increasing magnetic field. These oscillations are similar to the quantum oscillations in Shubnikov-de Haas \cite{brch82} and de Haas-van Alphen \cite{LL5,lai2001} effects.
	The dependence of the degenerate electron gas pressure along the magnetic field $P_{ez}$ on the electron number density for different magnetic fields is determined by the formula \cite{lai2001}

	\begin{equation}
	\begin{gathered}
	P_{ez} = \frac{1}{4\pi^2} \frac{eB}{c\hbar^2}\sum_{n = 0}^{\infty}g_n\int_{-\infty}^{+\infty} f_n \frac{p_z^2}{m_e} dp_z,
	\end{gathered}
	\label{nmd12}
	\end{equation}
	where $f_n$ is defined in \eqref{fd}. In the fully degenerate limit, to exclude the chemical potential, taking into account \eqref{nmd9}, \eqref{nmd10}, we have
	
	\begin{equation}
	\begin{gathered}
	P_{ez} =\frac{\sqrt{2m_e}}{3\pi^2}\frac{eB}{c\hbar^2}\sum_{n = 0}^{\frac{\mu_e}{\hbar \omega_B}} g_n(\mu_e-n\hbar \omega_B)^{3/2},
	\end{gathered}
	\label{nmd13}
	\end{equation}
	
	The transversal pressure calculation is similar to (\ref{nmd12}),(\ref{nmd13}), with the longitudinal momentum square changed by the transversal momentum square (\ref{tr_mom}). When computing the energy density, one should use formula (\ref{energy}).
	
	For a strong quantizing magnetic field, when all fully degenerate electrons occupy the ground Landau level, the longitudinal pressure, taking into account \eqref{nmd9},\eqref{nmd10}, is determined by the expression
	
	\begin{equation}
	\begin{gathered}
	P_{ez} =\frac{\sqrt{2m_e}}{3\pi^2}\frac{eB}{c\hbar^2}\mu_e^{3/2}=\frac{4\pi^4}{3}
	\frac{\hbar^6}{m_e^3}\frac{n_e^3}{(\hbar \omega_B)^2}.
	\end{gathered}
	\label{nmd14}
	\end{equation}
	It is seen that the pressure decreases with increasing field $\sim B^{-2}$ for a constant number density (see Fig. \ref{fig:fig5}). The interpretation of this result is related to the increase in the transversal phase volume of the ground level (\ref{phv}) with increasing field $\sim B$ and the corresponding increase in the permissible number of electrons at this level. The number of occupied phase space cells in the longitudinal direction correspondingly decreases at the same number density, thereby decreasing the longitudinal pressure and chemical potential (see Fig. \ref{fig:chempot}). Equation \eqref{nmd14} also suggests that the pressure increases as $\sim n_e^{3}$ for strong quantization in a given magnetic field.

	\subsection{Electron-positron pair creation in a magnetic field: nonrelativistic case}
	
	The creation of electron-positron pairs in a magnetic field has been considered by many authors \cite{shulman77,kamyak1992,kamyak1993, mush18}. The equilibrium electron and positron number densities at a given temperature $T$ and density $\rho$ are determined by the null sum of their chemical potentials and the quasi-neutrality condition (\cite{zelnov, bisn}):
	
	\begin{equation}
	\mu_{e+}^r + \mu_{e-}^r = 0, \quad \mu_{e+}^r =\mu_{e+}+m_e c^2 \quad \mu_{e-}^r=\mu_{e-}+m_e c^2, \qquad
	n_p = n_{e-} - n_{e+}.
	\label{paireq}
	\end{equation}
	Here, $n_p=\rho/m_p$ is the proton number density (we consider hydrogen), the signs $+$ and $-$ in the subscripts correspond to positrons and electrons, respectively, and the superscript $r$ in chemical potentials means their relativistic values. The particle number densities $n_{e\pm}$ are given by formulas \eqref{fd} and \eqref{relconsentr}. From the thermodynamic equilibrium of the $e+e-$-pairs \eqref{paireq}, it follows that
	
	\begin{equation}
	\mu_{e+} + \mu_{e-} = -2m_e c^2.
	\label{par1}
	\end{equation}
	In a nonquantizing magnetic field, the number density of $e^+e^-$-pairs in a nonrelativistic gas at $k_BT \ll m_ec^2$ is determined in the same way as without the magnetic field \cite{LL5}. From Eqn \eqref{nk2}, taking into account \eqref{paireq} and \eqref{par1}, for nondegenerate electrons, we obtain
	
	\begin{equation}
	\begin{gathered}
	n_{e+}n_{e-} = 4\bigg(\frac{m_ek_BT}{2\pi\hbar^2}\bigg)^{3}\exp\bigg(-\frac{2m_ec^2}{k_BT}\bigg).
	\end{gathered}
	\label{nepmnob}
	\end{equation}
	
	Taking into account the last relation in \eqref{paireq}, from \eqref{nepmnob}, we find \cite{LL5}
	
	\begin{equation}
	\begin{gathered}
	n_{e+}=n_{e-}-n_p =-\frac{n_p}{2}+\bigg[\frac{n_p^2}{4}+4\bigg(\frac{m_ek_BT}{2\pi\hbar^2}\bigg)^{3}
	\exp\bigg(-\frac{2m_ec^2}{k_BT}\bigg)\bigg]^{1/2}.
	\end{gathered}
	\label{par2}
	\end{equation}
	For nondegenerate, nonrelativistic electrons in an arbitrary magnetic field, from \eqref{nmd5}, taking into account \eqref{paireq} and \eqref{par1}, we have
	
	\begin{eqnarray}
	n_{e+}n_{e-}=
	\frac{m_e k_B T}{2\pi^3} \bigg(\frac{eB}{c\hbar^2}\bigg)^2\bigg\{\left[1-\exp(-\frac{eB \hbar}{m_ec k_B T})\right]^{-1}-\frac{1}{2}\bigg\}^2 \exp\bigg(-\frac{2m_ec^2}{k_BT}\bigg).
	\label{par3}
	\end{eqnarray}
	Taking into account the last relation in \eqref{paireq}, from \eqref{par3}, we obtain 
	
	\begin{equation}
	\begin{gathered}
	n_{e+}=n_{e-}-n_p =-\frac{n_p}{2}+\bigg[\frac{n_p^2}{4}+\frac{ m_e k_B T}{2\pi^3}
	\bigg(\frac{eB}{c\hbar^2}\bigg)^2\\
	\bigg\{\left[1-\exp\left(-\frac{eB \hbar}{m_ec k_B T}\right)\right]^{-1}
	-\frac{1}{2}\bigg\}^2 \exp\bigg(-\frac{2m_ec^2}{k_BT}\bigg)\bigg]^{1/2}.
	\end{gathered}
	\label{par4}
	\end{equation}
	The electron-positron pair number density in a non-degenerate, nonrelativistic gas in a quantizing magnetic field was considered in papers \cite{shulman77, kamyak1993}. In the limit $B = 0$, formula \eqref{par4} transforms into Eqn \eqref{par2} without a magnetic field, and in the case $\hbar \omega_B \gg k_BT$, the pair number density is $n_{e\pm} \sim B\sqrt{T}\exp{(-m_ec^2/k_BT)}$.
	
\begin{figure}[!htp]
		\centering
		\includegraphics[width=9.0cm,height=6.0cm]{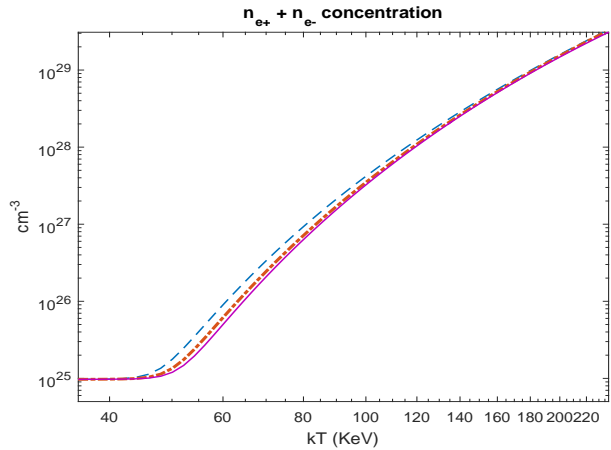}
		\caption{ Dependence of the pair number density $n_{e-} + n_{e+}$ in a nonrelativistic gas on the temperature in magnetic fields $B = 10^{12}; 10^{13}; 2\cdot10^{13}$ G (violet solid, red dashed-dotted, and blue dashed curves, respectively), with constant proton density $n_p = 10^{25}$ cm$^{-3}$. Clearly, the nonrelativistic magnetic field increases the pair production rate by increasing pair number density by not more than an order of magnitude compared to that in the field-free case. }
		\label{fig:fig7}
	\end{figure}

\begin{figure}[!htp]
		\centering
		\includegraphics[width=9.0cm,height=6.0cm]{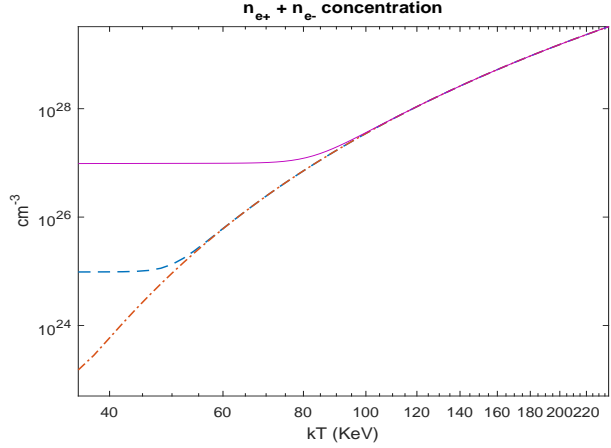}
		\caption{ Dependence of the pair number density $n_{e-} + n_{e+}$ in a nonrelativistic gas on the temperature in the magnetic field $B = 10^{13}$ G for proton number densities $n_p = 10^{23}; 10^{25}; 10^{27}$ cm$^{-3}$ (red dashed- dotted, blue dashed, and magenta curves, respectively). }
		\label{fig:fig8}
	\end{figure}

\begin{figure}[!htp]
		\centering
		\includegraphics[width=9.0cm,height=6.0cm]{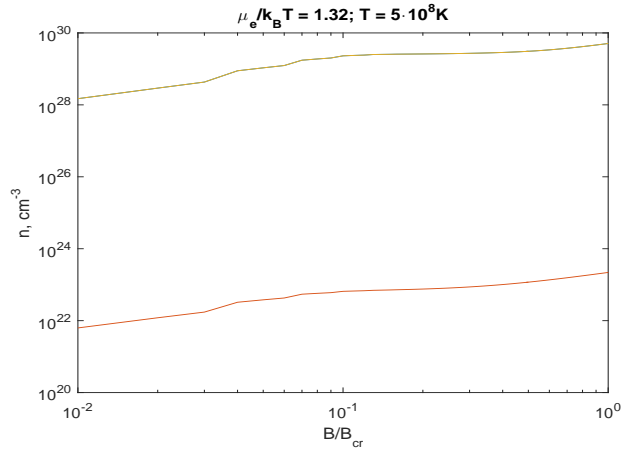}
		\caption{ Dependence of the number density of pairs $n_{e-} + n_{e+}$ (upper curve) and positrons $n_{e+}$ (bottom curve) in a nonrelativistic degenerate gas on the magnetic field at temperature $T = 5\cdot10^8$ K and a degeneracy parameter $\mu_e/k_BT = 1.32 \,\,\,(\mu_e/m_ec^2 = 0.2)$. }
		\label{fig:fig9}
	\end{figure}

	To find the number density of degenerate electrons and positrons in a magnetic field, in \eqref{par3}, one should use \eqref{relconsentr} and \eqref{fd} instead of \eqref{nmd5}; when deriving expressions for a number density like \eqref{par4}, the same substitution should be made. The resulting formulas turn out to be much more complicated than \eqref{par3} and \eqref{par4}.
	In a nondegenerate plasma with a given density $\rho=n_p m_p$, the pair creation in a magnetic field, according to \eqref{nepmnob}-\eqref{par4}, is determined by two dimensionless parameters: 
	
	\begin{equation}
	\frac{k_B T}{m_e c^2}, \quad \frac{\hbar\omega_B}{m_e c^2}.
	\label{par5}
	\end{equation}
	In a more general case, when taking into account the degeneracy according to \eqref{relconsentr} and \eqref{fd}, the dependence on the degeneracy parameter $\mu_e/k_B T$ appears, as in \eqref{nmd8}. The pair creation rate as a function of the magnetic field at different temperatures, densities, and degeneracy degrees is presented in Figs \ref{fig:fig7}-\ref{fig:fig9}.

	\section{Relativistic electron Fermi gas and $e^+e^-$-pairs in a supercritical magnetic field}
	
	\subsection{Fermi-Dirac distribution of relativistic electrons in an arbitrarily strong magnetic field}
	
	To calculate the number density of relativistic electrons, in equation \eqref{relconsentr} one should substitute the energy \eqref{energy} by its relativistic extension \eqref{relen}. The derivation in Section 3.1 suggests that the expression for the transversal phase space remains intact in the relativistic case. The transversal momentum \eqref{tr_mom} also remains valid for any magnetic field. The electron number density expressed in relativistic units takes the following form \cite{bag2002,canuto1968,lai2001}:
	
	\begin{equation}
	\begin{gathered}
	n_e = \frac{b}{4\pi^2} \frac{(m_ec)^3}{\hbar^3}\sum_{n = 0}^{\infty}g_n\int_{-\infty}^{+\infty}dp' f_n(p', \mu_{e}^{r},T),
	\end{gathered}
	\label{ne_final}
	\end{equation}
	where
	
	\begin{equation}
	\begin{gathered}
	f_n(p', \mu_{e}^{r},T) =\bigg[1 + \exp\bigg(\frac{m_ec^2\sqrt{1 + p'^2 + 2bn}- \mu_e^{r}}{k_BT}\bigg)\bigg]^{-1}, \quad p' = p_z/m_e c, \quad  \mu_e^r = \mu_e+m_e c^2.
	\end{gathered}
	\label{frel}
	\end{equation}
	The above considerations concerning the strongly quantizing and nonquantizing limits remain valid because, in the relativistic case, the electron transversal energy dependence on the magnetic field changes: for $b > 1$, the field turns out to be strongly quantizing if $ \sqrt{ m_ec^2\hbar\omega_B} > k_BT$. In the relativistic case, in a nonquantizing magnetic field, the summation in \eqref{ne_final},\eqref{frel} can be substituted by integration by analogy with \eqref{nk}. As in the nonrelativistic case, the number density of a relativistic Fermi gas of free electrons is found to be
	
	\begin{equation}
	\begin{gathered}
	n_e =  \frac{1}{4\pi^2} \frac{1}{\hbar^3}\int_{0}^{\infty}dp_\perp^2\int_{-\infty}^{+\infty}dp_z \bigg[1 + \exp\bigg(\frac{\sqrt{m_e^2c^4 + p_z^2 c^2+ p^{2}_{\perp}c^2}- \mu_e^{r}}{k_BT}\bigg)\bigg]^{-1}.
	\end{gathered}
	\label{woBrel}
	\end{equation}

	\subsection{Strongly degenerate electrons}
	
	In the shells of strongly magnetized neutron stars, electrons can be relativistic and strongly degenerate in a quantizing magnetic field [52], when $m_e c^2 \hbar \omega_B \sim (\mu_e^r)^2 \gg m_e^2 c^4$. For fully degenerate electrons, the integration in (4.1) is performed within the limits
	
	\begin{equation}
	\begin{gathered}
	-\sqrt{\bigg(\frac{\mu_e^r}{m_e c^2}\bigg)^2-1-2nb}\le\, p'\,\le
	\sqrt{\bigg(\frac{\mu_e^r}{m_e c^2}\bigg)^2-1-2nb}.
	\end{gathered}
	\label{grb6}
	\end{equation}
	Only terms with numbers $n\le [(\frac{\mu_e^r}{m_e c^2})^2-1]/2b$ remain in sum (4.1). In the absence of a magnetic field, the electron number density takes the form [36]
	
	\begin{equation}
	\begin{gathered}
	n_{ed}=\frac{(m_ec)^3}{3\pi^2\hbar^3}\bigg[\bigg(\frac{\mu_e^r}{m_e c^2}\bigg)^2-1\bigg]^{3/2}
	\end{gathered}
	\label{grb6b}
	\end{equation}
	
	In the relativistic fully degenerate case, the relative number density follows from (4.1), (4.5):
	
	\begin{equation}
	\begin{gathered}
	\frac{n_e}{n_{ed}}=\frac{3}{4}\frac{2b}{\mu_r^2 - 1}\sum_{n = 0}^{[(\mu_r^2-1)/2b]} g_n\sqrt{1-n\frac{2b}{\mu_r^2 - 1}}, \quad \mu_r=\frac{\mu_e^r}{m_e c^2}.
	\end{gathered}
	\label{grb6a}
	\end{equation}
	In the nonrelativistic limit $\mu_e^r=\mu_e+m_e c^2$, $\mu_e \ll m_e c^2$, the quantity $\frac{2b}{\mu_r^2 - 1}$  coincides with the parameter $\frac{\hbar \omega_B}{\mu_e}$ for a nonrelativistic gas \eqref{nmd9}. Its equality to unity means, that all electrons occupy only the ground Landau level. The dependence of the relative number density on $B/B_{cr}$ shown in Fig. 10, in which only relativistic effects are taken into account precisely, coincides qualitatively with that for the nonrelativistic limit shown in Fig. 3. For large $b$, only the term with $n = 0$ corresponding to the ground level remains. For
	
	\begin{equation}
	\begin{gathered}
	b_1=\bigg[\bigg(\frac{\mu_e^r}{m_e c^2}\bigg)^2-1\bigg]/4\le b\le \bigg[\bigg(\frac{\mu_e^r}{m_e c^2}\bigg)^2-1\bigg]/2=b_0
	\end{gathered}
	\label{grb7}
	\end{equation}
	he electron number density in the degenerate plasma is determined by the ground and first levels:
	
	\begin{figure}[!htp]
		\centering
		\includegraphics[width=9.0cm,height=6.0cm]{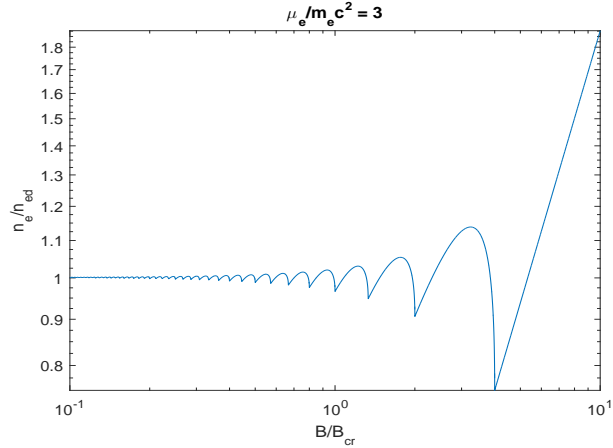}
		\caption{ Electron number density ratio in the case of quantizing and nonquantizing magnetic fields $\frac{n_e}{n_{ed}}$ as a function of the normalized magnetic field $B/B_{cr}$ in the case of relativistic and strongly degenerate electrons, according to \eqref{ne_rel} and \eqref{grb6}. The level of the proper relativism of electrons is determined by the chemical potential ${\mu_e^r}$ corresponding to the Fermi energy of fully degenerate electrons in the absence of a magnetic field. In this case, $\frac{\mu_e^r}{m_ec^2} = \mu_r = 3$. Here, the oscillations, as in the nonrelativistic case presented in Fig.\ref{fig:fig3}, are related to the quantizing magnetic field determining the discrete occupation of the low Landau levels. The transition of all electrons onto the ground level occurs, according to \eqref{grb7}, at $b=b_0=4$, corresponding to the last peak in this figure. This corresponds to $b = \big[\mu_r^2- 1\big]/2$. }
		\label{fig:deggas}
	\end{figure}
	
	\begin{equation}
	\begin{gathered}
	n_e = \frac{b}{2\pi^2}\frac{(m_ec)^3}{\hbar^3}
	\bigg[\sqrt{\bigg(\frac{\mu_e^r}{m_e c^2}\bigg)^2-1}
	+2\sqrt{\bigg(\frac{\mu_e^r}{m_e c^2}\bigg)^2-1-2b}\bigg]
	\quad
	{\mbox{at}}\quad
	b_1 \leq b \leq b_0\\
	n_e = \frac{b}{2\pi^2} \frac{(m_ec)^3}{\hbar^3}\sqrt{\bigg(\frac{\mu_e^r}{m_e c^2}\bigg)^2-1}
	\quad
	{\mbox{at}} \quad b \ge b_0.
	\end{gathered}
	\label{grb8}
	\end{equation}
	As in the nonrelativistic case \eqref{nmd11}, when the magnetic field crosses the boundary value, the transition down from the upper level is accompanied by a jump in the derivative of the function $n_e|_{\mu_e^r}(b)$ upon changing its sign (see Figs \ref{fig:fig3} and \ref{fig:deggas}).
	
	Let us find the longitudinal pressure of relativistic electrons $P_{ez}$ in a quantizing supercritical magnetic field. This pressure is expressed as a sum over the Landau levels, similar to the nonrelativistic case \eqref{nmd12}:
	
	\begin{equation}
	\begin{gathered}
	P_{ez}^r = \frac{1}{4\pi^2} \frac{eB}{c\hbar^2}\sum_{n = 0}^{\infty}g_n\int_{-\infty}^{+\infty} f_n p_z v_z dp_z,
	\end{gathered}
	\label{pez_rel0}
	\end{equation}	
	where \cite{LL2}
	\begin{equation}
	\begin{gathered}
	v_z = \frac{cp_z}{\sqrt{p_n^2 + m_e^2c^2}}.
	\end{gathered}
	\label{velz}
	\end{equation}
	Here, $f_n$ is defined in (4.2), $p_n^2 = p_z^2 + p_{\perp,n}^2 = p_z^2 + \frac{eB\hbar}{c}(2n + 1)$ (see (2.5)). Note that, in papers [22, 53], a somewhat different expression for the magnetized electron pressure was used, ignoring the dependence of the transversal momentum on the magnetic field (2.5). From (4.9) and (4.10), using the notation \eqref{frel}, we obtain the pressure of a magnetized relativistic electron gas in the form
	
	\begin{equation}
	\begin{gathered}
	P_{ez}^r = \frac{b}{4\pi^2} \frac{m_e^4c^5}{\hbar^3}\sum_{n = 0}^{\infty}g_n\int_{-\infty}^{+\infty}dp' f_n(p', \mu_{r},t) \frac{p'^2}{\sqrt{1 + p'^2 + 2bn + b}}.
	\end{gathered}
	\label{pez_rel1}
	\end{equation}
	
\begin{figure}[!htp]
		\centering
		\includegraphics[width=9.0cm,height=6cm]{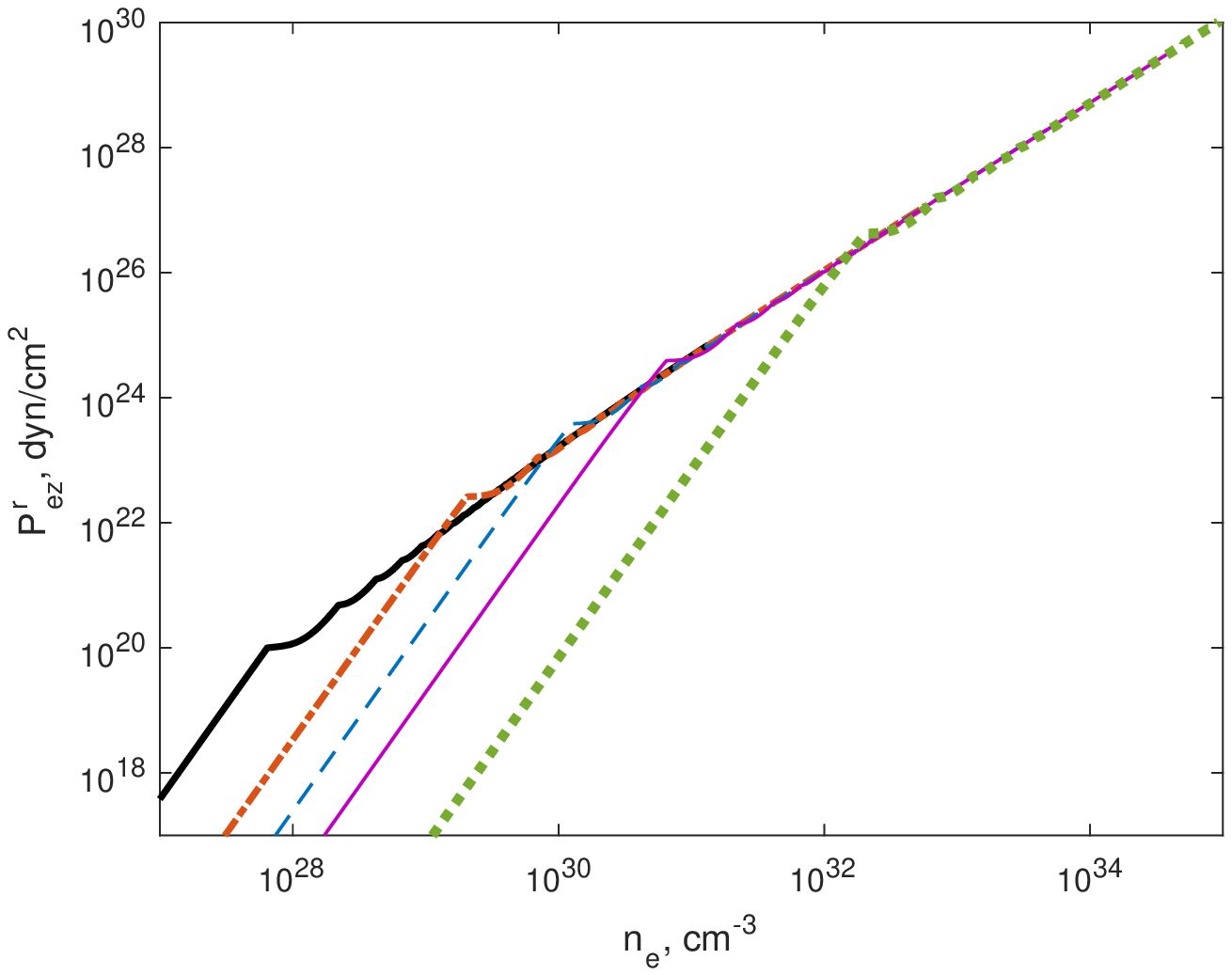}
		\includegraphics[width=9.0cm,height=6cm]{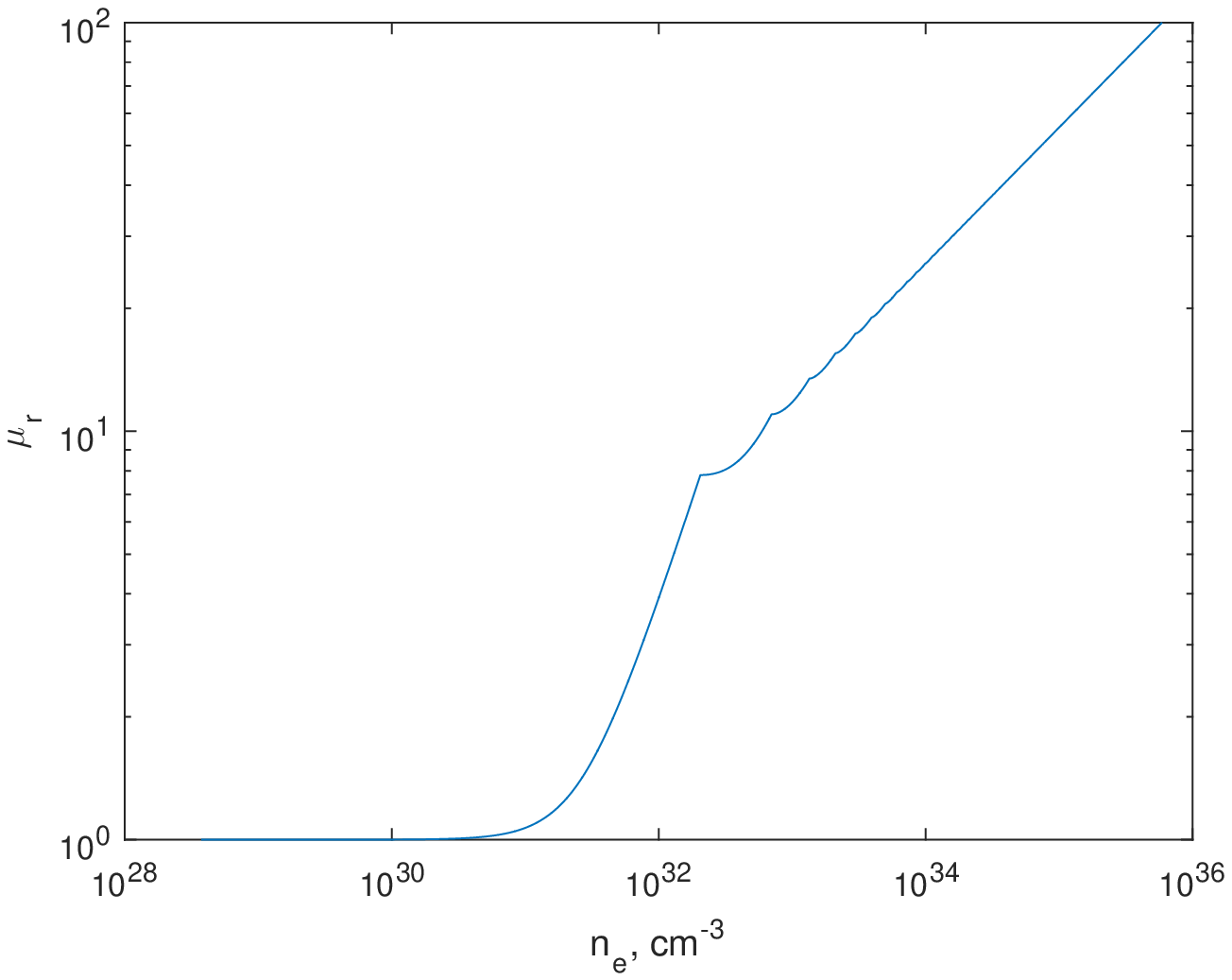}
		\caption{{\bf upper panel}: Pressure of a relativistic degenerate electron gas in a magnetic field as a function of the electron density in the magnetic fields $B/B_{cr} = 0.03; 0.3; 1; 3; 30$ (black solid, red dashed-dotted, blue dashed, magenta, and green dashed curves, respectively). The black curve in the nonrelativistic region coincides with the similar curve shown in Fig. \ref{fig:fig4}. When the electron density increases in a constant field, the pressure tends to the adiabatic law with ultra-relativistic index $\gamma = 4/3$.\newline {\bf lower panel}: Electron gas chemical potential  $\mu_r = \mu_e^r/m_ec^2$ in a magnetic field as a function of the density ne in a strong field, $B/B_{cr} = 30$. The chemical potential exceeds unity $(\mu_r>1)$ near the beginning of electron transitions up to higher Landau levels, where, in \eqref{pez0rel}, the parameter $y\sim 1$. In the transition region, the dependence of pressure on density changes starting from $P_{ez} \sim n_e^3$, corresponding to all electrons occupying the ground level, to $P_{ez} \sim n_e^{4/3}$, corresponding to an ultra-relativistic electron gas in a nonquantizing magnetic field.	}
		\label{fig:relP1}
	\end{figure}

\begin{figure}[!htp]
		\centering
		\includegraphics[width=9.0cm,height=6.0cm]{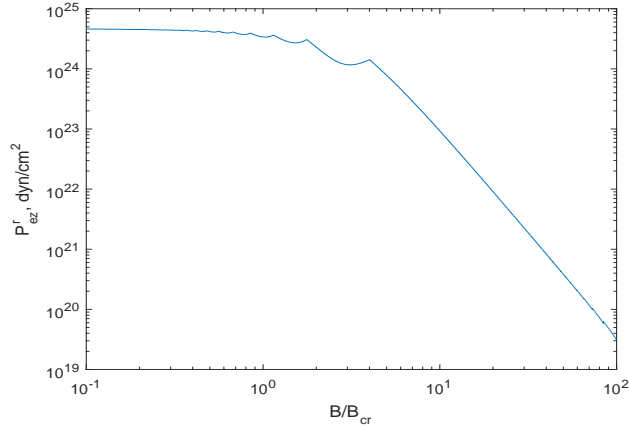}
		\caption{ Degenerate electron gas pressure as a function of $B/B_{cr}$ for an electron density of $n_e = 1\cdot10^{31}$ cm$^{-3}$. The pressure oscillations in a relativistic gas are similar to those in the nonrelativistic case. With increasing magnetic field, an increasingly fewer number of Landau levels are populated by electrons. After the last pressure maximum, all electrons transit onto the ground Landau level, and $P_{ez}^r \propto 1/B^2\sqrt{B}$.}
		\label{fig:relP2}
	\end{figure}

	In the strongly degenerate limit $T = 0$, formula (4.11) takes the form
	\begin{equation}
	\begin{gathered}
	P_{ez}^r = \frac{b}{2\pi^2} \frac{m_e^4c^5}{\hbar^3}\sum_{n = 0}^{[(\mu_r^2-1)/2b]}g_n\int_{0}^{\sqrt{\mu_r^2-1-2bn}}dp' \frac{p'^2}{\sqrt{1 + p'^2 + 2bn + b}}.
	\end{gathered}
	\label{pez_rel2}
	\end{equation}
	The integral in \eqref{pez_rel2} is calculated analytically using the change of variables and the table integrals \cite{integ}:
	\begin{equation}
	\begin{gathered}
	z = \arsh \frac{p'}{\sqrt{1 + 2bn + b}},\;\;\;  dz = \frac{dp'}{\sqrt{1 + p'^2+ 2bn + b}}, \;\;\; \sinh z =  \frac{p'}{\sqrt{1 + 2bn + b}} \\
	\int \frac{p'^2}{\sqrt{1 + p'^2+ 2bn + b}}dp' = (1 + 2bn + b)\int (\sinh^2z) dz, \;\;\; \int (\sinh^2z) dz = \frac{1}{4}\big[\sinh(2z) - 2z\big], \\ \sinh 2z= 2 \sinh z \cosh z = 2 \sinh z\sqrt{1 + \sinh^2z}, \;\;\;\arsh x = \log(\sqrt{x^2 + 1} + x).
	\end{gathered}
	\label{intgiperb}
	\end{equation}
	Using \eqref{velz}, from \eqref{pez_rel2}, we obtain the following expression for the pressure of a degenerate gas with any relativism degree in an arbitrary magnetic field: 
	
	\begin{equation}
	\begin{gathered}
	P_{ez}^r = \frac{b}{4\pi^2} \frac{m_e^4c^5}{\hbar^3}\sum_{n = 0}^{[(\mu_r^2-1)/2b]}g_n(1 + 2bn + b)\big[x\sqrt{x^2 + 1} - \log(x + \sqrt{x^2 + 1})\big]; \\ x = \sqrt{\frac{\mu_r^2 - 1 - 2bn}{1 + 2bn + b}}.
	\end{gathered}
	\label{pez_rel}
	\end{equation}
	Figures \ref{fig:relP1} and \ref{fig:relP2} present the dependence of a degenerate electron gas on the density and magnetic field. In the limit $\mu_r^2 - 1 \ll 1$ and $b \ll 1$, formula (4.14) transforms to the nonrelativistic limit\footnote{When considering the nonrelativistic limit $z \ll
		1$, we use the expansion $\frac{1}{4}\big[\sinh (2z) - 2z\big]\approx \frac{z^3}{3}$.} (3.27).

	Let us consider now a strongly quantizing magnetic field $b > \frac{\mu_r^2 - 1}{2}$. In this case, in formula \eqref{pez_rel} the summation over the Landau level is not needed; only the term with $n = 0$
	
	\begin{equation}
	\begin{gathered}
	P_{ez}^r = \frac{b}{4\pi^2} \frac{m_e^4c^5}{\hbar^3}(1 + b)\bigg[\frac{\sqrt{(\mu_r^2 + b)(\mu_r^2-1)}}{1 + b} - \log(\sqrt{\frac{\mu_r^2 - 1}{1 + b}} +\sqrt{\frac{\mu_r^2 + b}{1 + b})}\bigg].
	\end{gathered}
	\label{pez3}
	\end{equation}
	Taking into account the number density \eqref{grb6}, \eqref{grb6a}, this formula takes the form
	\begin{equation}
	\begin{gathered}
	P_{ez}^r = \frac{b}{4\pi^2} \frac{m_e^4c^5}{\hbar^3} (1 + b) \big[y\sqrt{y^2 + 1} - \log(y + \sqrt{y^2 + 1})\big]; \\ y = \frac{2\pi^2\hbar^3}{m_e^3c^3}\frac{n_e}{b\sqrt{1 + b}}.
	\end{gathered}
	\label{pez0rel}
	\end{equation}
	With increasing magnetic field, the parameter $y\equiv x$ from \eqref{pez_rel} and the quantity \eqref{pez0rel} become much smaller than one, as in the nonrelativistic case. Using the same expansion, we obtain the longitudinal pressure in the form
	
	\begin{equation}
	\begin{gathered}
	P_{ez}^r = \frac{4\pi^4}{3} \frac{\hbar^6}{m_e^5c^4}\frac{n_e^3}{b^2\sqrt{1 + b}}.
	\end{gathered}
	\label{pez4}
	\end{equation}
	Thus, in the limit of a super-strong field, when all electrons occupy the ground Landau level, the pressure $P_{ez}^r \propto n_e^3/B^2\sqrt{B}$. Note that electrons concentrate at the ground level, according to \eqref{pez_rel2}, only if $\mu_r \le \sqrt{1+2b}$, i.e., if $y \le \sqrt{\frac{2b}{1+b}}$. Otherwise, the occupation of upper levels occurs, and for $\mu_r \gg b$, the transition to the quasi-classical limit with a nonquantizing magnetic field happens. In a strongly quantizing magnetic field, the dependence of the longitudinal pressure on the electron number density is the same for any relativism degree (but a narrow region with $y \sim 1$ near the transition to the nonquantizing case). It has a universal character, $P_{ez} \sim n_e^3$, unlike in a nonquantizing field, where the dependence of pressure on the electron number density changes from 5/3 to 4/3 when passing from nonrelativistic to relativistic electrons.

	\subsection{Electron-positron pair creation in a magnetic field: relativistic case}
	
	In the case of relativistic electrons in an arbitrary magnetic field, the equilibrium number density of electron-positron pairs, as in the nonrelativistic case, is determined by formula \eqref{paireq}. To obtain the number density of degenerate relativistic electrons and positrons in a magnetic field, in \eqref{par3}, we should use \eqref{ne_final} and \eqref{frel} instead of \eqref{nmd5}. The same substitution should be made when deriving relations for a number density like in \eqref{par4}.
	
	Simple expressions for the number density of pairs can be found in the ultrarelativistic limit in a nonquantizing field $k_B T\gg m_e c^2$, $ k_BT\gg \sqrt{ m_ec^2\hbar\omega_B}$. In this case, the magnetic
	field does not affect pair creation. From \eqref{woBrel}, in the ultrarelativistic limit, we find
	
	\begin{equation}
	\begin{gathered}
	n_e =  \frac{1}{4\pi^2} \frac{1}{\hbar^3}\int_{0}^{\infty}dp_\perp^2\int_{-\infty}^{+\infty}dp_z \bigg[1 + \exp\bigg(\frac{\sqrt{p_z^2 c^2+ p^{2}_{\perp}c^2}- \mu_e^{r}}{k_BT}\bigg)\bigg]^{-1}\\=\frac{1}{2\pi^2 \hbar^3}\int_{0}^{\infty}\bigg[1 + \exp\bigg(\frac{pc- \mu_e^{r}}{k_BT}\bigg)\bigg]^{-1}p^2 dp,\quad p^2
	=p_z^2 c^2+ p^{2}_{\perp}c^2.
	\end{gathered}
	\label{urel}
	\end{equation}
	By designating $x=\frac{pc}{k_B T}$, $\xi=\frac{\mu_e^r}{kT}$, we get the electron number density
	
	\begin{equation}
	\begin{gathered}
	n_{e^-} = \frac{1}{2\pi^2}\bigg(\frac{k_B T}{c \hbar}\bigg)^3\int_{0}^{\infty}\frac{x^2 dx}{1+\exp(x-\xi)}
	= \frac{1}{2\pi^2}\bigg(\frac{k_B T}{c \hbar}\bigg)^3F_2(\xi).
	\end{gathered}
	\label{urel1}
	\end{equation}
	The Fermi function $F_2(\xi)$ is defined in \eqref{nkf}. The positron number density taking into account \eqref{paireq} reads
	
	\begin{equation}
	\begin{gathered}
	n_{e^+} = \frac{1}{2\pi^2}\bigg(\frac{k_B T}{c \hbar}\bigg)^3\int_{0}^{\infty}\frac{x^2 dx}{1+\exp(x+\xi)}
	= \frac{1}{2\pi^2}\bigg(\frac{k_B T}{c \hbar}\bigg)^3F_2(-\xi).
	\end{gathered}
	\label{urel2}
	\end{equation}
	When the number density of pairs far exceeds that of original electrons, namely $\mu_e=\xi \simeq 0$, and the sum of electron and positron number densities is expressed as \cite{nad,zelnov,bisn}
	
	\begin{equation}
	\begin{gathered}
	n_{e+} + n_{e-} \simeq \frac{1.803}{\pi^2}\bigg(\frac{k_B T}{c\hbar}\bigg)^3,\quad F_2(0)=\frac{3}{2}\xi(3)=1.803,
	\end{gathered}
	\label{nepmnobrel0}
	\end{equation}
	where $\xi(x)$ is Riemann’s $\xi$-function.
	
	\indent\indent Of particular interest is the case where electrons move nonrelativistically along the field, which occurs at  $k_B T \ll m_e c^2$, in a supercritical magnetic field $B>B_{cr}$ (see \eqref{enb}). Such a situation can take place in the accretion columns of strongly magnetized neutron stars \cite{basko76,wf81,mstp15}. There, the magnetic field near the neutron star surface could be as high as $B\sim 10^{15}$ G with a proton number density of $n_p \sim 10^{25-27}$cm$^{-3}$ cm-3 and a temperature of $T\sim 10^9$ K. As shown in papers \cite{shulman77,mush18}, the presence of the magnetic field enhances the electron-positron pair production under the same $n_p, T$. In that case, the number density of $e^+e^-$ pairs can be calculated analytically.
	
	Let us present the electron number density \eqref{ne_final},\eqref{frel} in the following form:
	
	\begin{equation}
	\begin{gathered}
	n_e = \frac{b}{4\pi^2} \frac{(m_ec)^3}{\hbar^3}\sum_{n = 0}^{\infty}g_s\int_{-\infty}^{+\infty}dp'
	\bigg[1 + \exp\bigg(\frac{m_ec^2\sqrt{1 + p'^2 + 2bn}- \mu_e^{r}}{k_BT}\bigg)\bigg]^{-1}\\
	= \frac{b}{4\pi^2} \frac{(m_ec)^3}{\hbar^3}
	\bigg\{\int_{-\infty}^{+\infty}dp'
	\bigg[1 + \exp\bigg(\frac{m_ec^2\sqrt{1 + p'^2}- \mu_e^{r}}{k_BT}\bigg)\bigg]^{-1}\\+2\sum_{n = 1}^{\infty}\int_{-\infty}^{+\infty}dp'
	\bigg[1 + \exp\bigg(\frac{m_ec^2\sqrt{1 + p'^2 + 2bn}- \mu_e^{r}}{k_BT}\bigg)\bigg]^{-1}\bigg\}.
	\end{gathered}
	\label{ne_rel}
	\end{equation}
	Consider the limiting case $b \gg 1$. The formula above reduces to:
	
	\begin{equation}
	\begin{gathered}
	n_e = \frac{b}{4\pi^2} \frac{(m_ec)^3}{\hbar^3}
	\bigg\{\int_{-\infty}^{+\infty}dp'
	\bigg[1 + \exp\bigg(\frac{m_ec^2\sqrt{1 + p'^2}- \mu_e^{r}}{k_BT}\bigg)\bigg]^{-1}\\+2\sum_{n = 1}^{\infty}\int_{-\infty}^{+\infty}dp'
	\bigg[1 + \exp\bigg(\frac{m_ec^2\sqrt{p'^2 + 2bn}- \mu_e^{r}}{k_BT}\bigg)\bigg]^{-1}\bigg\}.
	\end{gathered}
	\label{grb}
	\end{equation}
	For the above plasma parameters in the accretion column, the degeneracy is insignificant, and the term with an exponent can be much larger than unity; therefore, we can use the Maxwell-Boltzmann distribution function, for which
	
	\begin{equation}
	\begin{gathered}
	n_e = \frac{b}{4\pi^2} \frac{(m_ec)^3}{\hbar^3}
	\bigg\{\int_{-\infty}^{+\infty}dp'
	\exp\bigg(-\frac{m_ec^2\sqrt{1 + p'^2}- \mu_e^{r}}{k_BT}\bigg)\\+2\sum_{n = 1}^{\infty}\int_{-\infty}^{+\infty}dp'
	\exp\bigg(-\frac{m_ec^2\sqrt{p'^2 + 2bn}- \mu_e^{r}}{k_BT}\bigg)\bigg\}.
	\end{gathered}
	\label{grb1}
	\end{equation}
	For nonrelativistic longitudinal velocities and supercritical magnetic fields, we can expand $\sqrt{1 + p'^2}=1+\frac{p'^2}{2}$ and $\sqrt{p'^2 + 2bn}=\sqrt{2bn}(1+ \frac{p'^2 }{2nb})$. By taking these expansions into account, Eqn \eqref{grb1} can be rewritten in the form
	
	\begin{equation}
	\begin{gathered}
	n_e = \frac{b}{4\pi^2} \frac{(m_ec)^3}{\hbar^3}
	\bigg\{\int_{-\infty}^{+\infty}dp'
	\exp\bigg(-\frac{m_ec^2(1 + p'^2/2)- \mu_e^{r}}{k_B T}\bigg)\\+2\sum_{n = 1}^{\infty}\int_{-\infty}^{+\infty}dp'
	\exp\bigg(-\frac{m_ec^2\sqrt{2bn}(1+p'^2/2nb)- \mu_e^{r}}{k_B T}\bigg)\bigg\}.
	\end{gathered}
	\label{grb2}
	\end{equation}
	
	By integrating over longitudinal momenta $dp'$, from \eqref{grb2} and taking into account \eqref{paireq}, we find for the number densities of electrons and positrons
	\begin{equation}
	\begin{gathered}
	n_{e-} = \frac{b}{4\pi^2} \frac{(m_ec)^3}{\hbar^3}\sqrt{\frac{2\pi k_B T}{m_e c^2}}
	\exp\bigg(\frac{\mu_e^r}{k_B T}\bigg)\bigg[\exp\bigg(-\frac{m_e c^2}{k_B T}\bigg)\\
	+ 2\sum_{n = 1}^{\infty}(2bn)^{1/4}
	\exp\bigg({-\frac{m_e c^2\sqrt{2bn}}{k_B T}}\bigg) \bigg],
	\\
	n_{e+} = \frac{b}{4\pi^2} \frac{(m_ec)^3}{\hbar^3}\sqrt{\frac{2\pi k_B T}{m_e c^2}}
	\exp\bigg(-\frac{\mu_e^r}{k_B T}\bigg)\bigg[\exp\bigg(-\frac{m_e c^2}{k_B T}\bigg)\\
	+ 2\sum_{n = 1}^{\infty}(2bn)^{1/4}
	\exp\bigg({-\frac{m_e c^2\sqrt{2bn}}{k_B T}}\bigg) \bigg],
	\end{gathered}
	\label{ne_rel3}
	\end{equation}
	From \eqref{ne_rel3}, in a similar way to the derivation of \eqref{nepmnob},\eqref{par3}, we obtain
	
	\begin{equation}
	\begin{gathered}
	n_{e-}n_{e+} = \frac{b^2}{8\pi^3} \frac{m_e^5c^4 k_B T}{\hbar^6}
	\bigg[\exp\bigg(-\frac{m_e c^2}{k_B T}\bigg) + 2\sum_{n = 1}^{\infty}(2bn)^{1/4}
	\exp\bigg({-\frac{m_e c^2\sqrt{2bn}}{k_B T}}\bigg) \bigg]^2,
	\end{gathered}
	\label{grb3}
	\end{equation}
	From \eqref{grb3} and \eqref{paireq}, by solving a quadratic equation, we find
	
	\begin{equation}
	\begin{gathered}
	n_{e+} = -\frac{n_p}{2}+\sqrt{\frac{n_p^2}{4} + \frac{ b^2 m_e^5c^4 k_B T}{8\pi^3\hbar^6}
		\bigg[\exp\bigg(-\frac{m_e c^2}{k_B T}\bigg) + 2\sum_{n = 1}^{\infty}(2bn)^{1/4}
		\exp\bigg({-\frac{m_e c^2\sqrt{2bn}}{k_B T}}\bigg) \bigg]^2}.
	\end{gathered}
	\label{grb4}
	\end{equation}
	For $B\gg B_{cr}$, most electrons occupy the ground level, so that the first term in square brackets in \eqref{grb4} is much larger than the whole sum, whose terms also rapidly decrease. By retaining the first term in the sum in \eqref{grb4}, we obtain

	\begin{equation}
	\begin{gathered}
	n_{e+}=-\frac{n_p}{2}+\sqrt{\frac{n_p^2}{4}+\frac{ b^2 m_e^5c^4 k_B T}{8\pi^3\hbar^6}
		\bigg[\exp\bigg(-\frac{m_e c^2}{k_B T}\bigg) + 2(2b)^{1/4}
		\exp\bigg({-\frac{m_e c^2\sqrt{2b}}{k_B T}}\bigg) \bigg]^2}.
	\end{gathered}
	\label{grb5}
	\end{equation}

\begin{figure}[!htp]
		\centering
		\includegraphics[width=9.0cm,height=6.0cm]{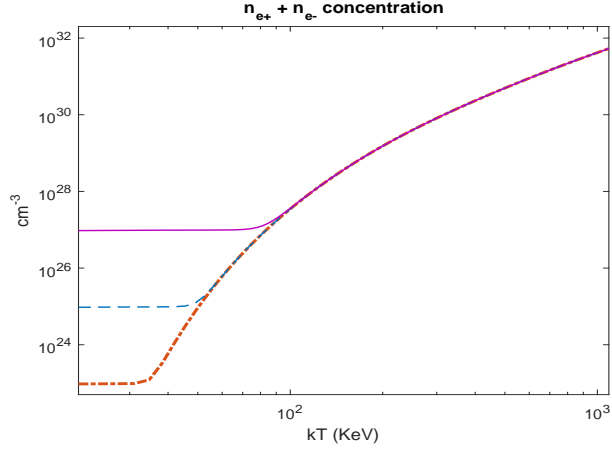}
		\caption{ Number density of pairs $n_{e-} + n_{e+}$ versus temperature in the magnetic field B= 1013 G for the proton number densities $n_p = 10^{23}; 10^{25}; 10^{27}$ cm$^{-3}$ (red dashed-dotted, blue dashed, and magenta curves, respectively). }
		\label{fig:pairs1}
	\end{figure}

\begin{figure}[!htp]
		\centering
		\includegraphics[width=9.0cm,height=6.0cm]{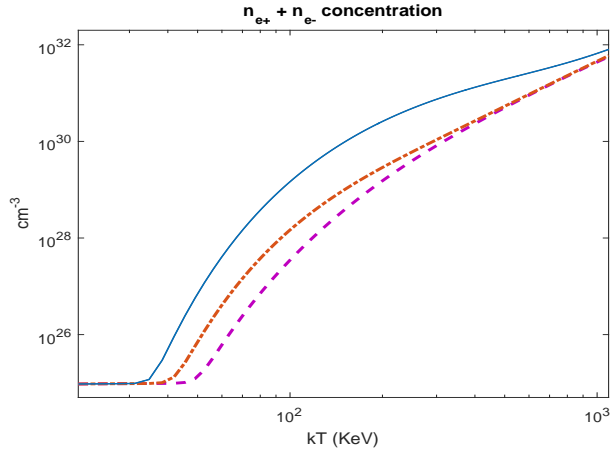}
		\caption{ Pair number density $n_{e-} + n_{e+}$ as a function of the temperature in the magnetic fields$B = 10^{13}; 10^{14}; 10^{15}$ G (magenta dashed, red dashed-dotted, and blue curves, respectively) at a constant proton density of $n_p = 10^{25}$ cm$^{-3}$. Clearly, with an increasing magnetic field, a significant pair creation rate occurs, increasing the pair number density. For other proton number densities, the dependences are qualitatively similar. }
		\label{fig:pairs2}
	\end{figure}

\begin{figure}[!htp]
		\centering
		\includegraphics[width=9.0cm,height=6.0cm]{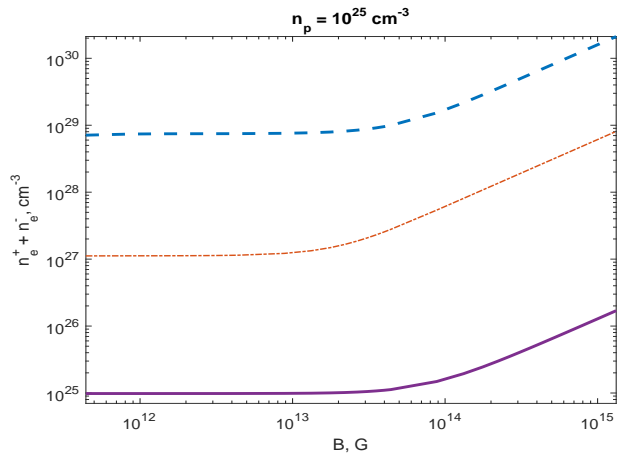}
		\caption{ Pair number density $n_{e-} + n_{e+}$ as a function of magnetic field $B$ at temperatures $T = 5\cdot 10^8; 1\cdot10^9; 2\cdot10^9K$ (magenta solid, red dashed-dotted, and blue dashed curves) at a constant proton number density of $n_p = 10^{25}$ cm$^{-3}$. The transition from weak dependence on the field, where many Landau levels are occupied, to a linear function in the strongly quantum case is visible. For other proton number densities, the dependences are qualitatively similar.}
		\label{fig:pairs3}
	\end{figure}

	Figures \ref{fig:pairs1} and \ref{fig:pairs2} present the number density of pairs as a function of the temperature for various magnetic fields and proton densities. Figure \ref{fig:pairs3} displays the electron-positron pair number density as a function of the magnetic field $B$ for a fixed proton density. For $m_ec^2\sqrt{b}/k_BT \gg 1$, corresponding to the occupation of the ground Landau level only, with increasing magnetic field, the dependence tends to linear, in agreement with \eqref{grb5}. A similar dependence takes place in the nonrelativistic limit (see \eqref{par4}).
	
	\section{Conclusion}
	
	We have considered the thermodynamic properties of an electron gas in an arbitrary magnetic field and any relativism degree of electrons. We have obtained analytical formulas in the limiting cases of the weak and strongly quantizing magnetic field. In agreement with the results of papers [71, 82], we have shown that, in a strong magnetic field, the pair production efficiency increases. This is due to a linear increase in the part of the phase volume transversal to the magnetic field B. We have investigated the properties of electrons in a quantizing magnetic field and oscillations of the thermodynamic functions in strong magnetic fields caused by discrete transitions between the Landau levels. We have found that, for a strongly degenerate gas in a strongly quantizing magnetic field, when all electrons occupy the ground Landau level, the electron pressure along the magnetic field increases as $P_{ez} \sim n_e^3$. At the same time, after all electrons transit onto the ground Landau level, the longitudinal electron pressure decreases as $P_{ez} \sim B^{-2}$. This is also a consequence of the increase in the transversal part of the phase volume noted above. In a relativistic gas, in the strongly quantizing limit $B/B_{cr} \gg 1$, pressure $P_{ez}^r \sim n_e^3/B^2\sqrt{B}$. Note that the dependence $P_{ez}^r \sim n_e^3$ takes place in both the relativistic and non-relativistic cases.

	{}

\end{document}